\documentclass[twocolumn]{aastex7}
\usepackage[utf8]{inputenc}
\usepackage{rotating}
\usepackage{graphicx}
\usepackage{amssymb}
\usepackage{amsmath}
\usepackage{hyperref}

\usepackage{txfonts}

\usepackage{color}
\usepackage{comment}
\usepackage{amsmath,bm}
\usepackage{siunitx}


\newcommand{\be}{\begin{equation}}
\newcommand{\ee}{\end{equation}}
\newcommand{\Edot}{\dot{E}}

\begin{document}
\defcitealias{Kaplan2011}{K+11}

\title{
UVOIR spectrum,  X-ray emission, and proper motion 
of the isolated neutron star 
RX\,J2143.0+0654\footnote{Based on observations made with the NASA/ESA Hubble Space Telescope, obtained at the Space Telescope Science Institute, which is operated by the Association of Universities for Research in Astronomy, Inc., under NASA contract NAS 5-26555. These observations are associated with program \#17476.}
}

\author[0000-0002-7481-5259]{George G. Pavlov}
\affiliation{Department of Astronomy and Astrophysics, The Pennsylvania State University, 525 Davey Lab, University Park, PA 16802, USA}
\email{ggp1@psu.edu}

\author[0000-0002-9892-7546]{Vadim Abramkin}
\affiliation{Independent Researcher}
\email{vadab2077@gmail.com}

\author[0000-0003-2317-9747]{B. Posselt}
\affiliation{Oxford Astrophysics, University of Oxford,
Denys Wilkinson Building, Keble Road, 
Oxford OX1 3RH, UK
}
\affiliation{Department of Astronomy and Astrophysics, The Pennsylvania State University, 525 Davey Lab, University Park, PA 16802, USA}
\email{bettina.posselt@physics.ox.ac.uk}

\correspondingauthor{George Pavlov}
\email{ggp1@psu.edu}

\begin{abstract}
We observed the isolated neutron star RX\,J2143.0+0654 
with the Hubble Space Telescope (HST)
in the UVOIR wavelength range (0.14--1.7 $\mu$m).  
The UV part 
is consistent with
a Rayleigh-Jeans tail of a thermal spectrum, $f_\nu\propto \nu^2$, 
while a power-law spectrum, $f_\nu \propto \nu^\alpha$ with $\alpha \sim -0.8$,
dominates in the NIR-optical.
A joint fit of the UVOIR and contemporaneous X-ray spectra with a two-component blackbody with possible absorption features + power-law optical spectrum yields the following temperature and apparent radius of the colder component (which gives the main contribution in the UV): $kT_{\rm cold}\approx 45$ eV, $R_{\rm cold}\approx 6 d_{260}$ km, where $d_{260}$ is the distance in units of 260 pc.
The temperature and radius of the hotter component, $kT_{\rm hot}\approx 106$ eV and $R_{\rm hot} \approx 1.5d_{260}$ km, the parameters of an absorption feature at 0.74 keV, 
and the properties of X-ray pulsations, are the same as found in previous X-ray observations. 
In the NIR images the neutron star is possibly surrounded by extended emission with a characteristic size of $\sim 2''$ and flux densities of about 1.7 and 0.9 $\mu$Jy at 1.54 and 1.15 $\mu$m, respectively.
Comparison with a previous HST observation in the optical 14 years ago  
shows a proper motion 
$\mu\approx 6$ mas\,yr$^{-1}$, which corresponds to a small transverse velocity of 
$7d_{260}$ km\,s$^{-1}$. It is consistent with the hypothesis that the neutron star was born in the vicinity of the solar system about 0.5 Myr ago.
\end{abstract}

\section{Introduction}
\label{sec:intro}
\subsection{Properties of XTINSs}
\label{sec:xtins}
A family of 7 nearby ($d \lesssim 1$ kpc) neutron stars (NSs), 
very different from
commonly known rotation powered pulsars (RPPs), was discovered in 1990s with the X-ray observatory ROSAT (see 
\citealt{Haberl2007} and \citealt{Kaplan2009} for reviews). These young ($\lesssim 1$ Myr), slowly rotating
($P = 3$--12 s)
NSs show predominantly thermal X-ray spectra, with temperatures 
$kT \sim 50$--100 eV. Unlike the RPPs, their X-ray luminosities, $L_X\sim 10^{30.8}$--$10^{32.5}$ erg s$^{-1}$, are 
higher than the NS spin-down energy loss rates, $\dot{E}\sim 10^{29.4}$--$10^{30.7}$ erg s$^{-1}$, and, unlike RPPs, they do not show easily detectable nonthermal X-ray,
radio or $\gamma$-ray emission.
Their surface magnetic fields $B\sim 10^{13}$ G, as well as  positions in the period - period derivative ($P$-$\dot{P}$) diagram,
place them between ordinary RPPs and magnetars.
Various names and abbreviations have been suggested for this NS group (INS, RQINS, XTINS, XDINS, Mag7, etc). We will call them  XTINSs -- {\em X-ray Thermal Isolated Neutron Stars}.
Although only 7 XTINSs have been firmly detected (and a few candidates found with eROSITA -- 
\citealt{Pires2022}, \citealt{Kurpas2024}), 
population studies show that 
such NSs,  whose soft X-ray emission is absorbed in the ISM at large distances, may represent nearly
half of the local NS population (e.g., \citealt{Posselt2008}).

Fits of the X-ray spectra of XTINSs with blackbody (BB) models made it possible to estimate their surface temperatures $T$ and radius-to-distance ratios $R/d$ 
(\citealt{Haberl2007}, \citealt{Kaplan2009}). 
Some XTINSs have possibly shown 
variations of X-ray emission 
(\citealt{Haberl2006}, \citealt{Pires2019,Pires2023}),
but only for RX\,J0720.4--3125 there are enough data to test possible explanations (e.g., \citealt{Hohle2012}).
According to 
\citet{vanKerkwijk2007}, the variations cannot be explained by the previously proposed NS precession, while episodes of
accretion of circumstellar bodies, e.g., asteroids, remain a viable interpretation.

Based on the X-ray results, it was expected that the UV-optical (UVO) spectra of XTINSs are simply Rayleigh-Jeans (R-J) tails of the thermal X-ray spectra.
However,  Hubble Space Telescope (HST) observations of XTINSs
showed a more complicated picture. For instance, although the UVO spectrum of the brightest of these objects, RX\,J1856.5--3754 (J1856 hereafter) 
has an expected slope,
$f_\nu\propto\nu^2$, 
the UVO flux density $f_\nu$ is a factor of 7 above the extrapolation of the X-ray BB spectrum into this frequency range (e.g., 
\citealt{vanKerkwijk2001}).
Noticed also in other XTINSs,  such
{\em optical excess}
could be explained by a model in which X-rays come from a smaller, hotter area of the NS surface, created by anisotropic heat transfer in the strong magnetic field, while the bulk of the NS surface is responsible for the UVO emission (e.g., \citealt{Braje2002}).
An alternative explanation is that the optical excess is due to the difference of the assumed BB emission from those of actual NS atmospheres or solid surfaces
\citep{Pavlov1996,Perez-Azorin2006,Ho2007}. 

From HST observations of 5 other XTINSs in two bands, 
1380--1570 \AA\ and  
4000--5600 \AA, 
\citet{Kaplan2011} 
(K+11 
hereafter) found that not only  
the flux density  
values were above the continuation of the X-ray BB  into the UVO range, but also
{\em the spectral slopes were inconsistent with an R-J spectrum} 
for 4 of the 5 XTINSs.  If one naively connects the two spectral points with a straight line in the $\log\nu$-$\log f_\nu$ plane (i.e., assume $f_\nu \propto \nu^\alpha$),
then the slope 
varied from  $\alpha=1.63\pm 0.14$ to $\alpha=0.53\pm 0.08$,
instead of $\alpha=2$ for the R-J spectrum.  The reason for this unexpected behavior remains unknown. 

\citetalias{Kaplan2011} discussed hypothetical explanations,
such as nonthermal 
emission of relativistic electrons 
or resonant scattering of thermal photons in the NS magnetosphere.
\citet{Ertan2017} suggested that the optical emission is produced in an inner rim of a fallback disk heated by the X-ray irradiation and magnetic stresses.
Since 
different models predict different spectra, 
a comparison with observed spectra could allow one to choose the right model, but it is impossible with just two spectral points available.
Obviously,  more spectral points in a broader wavelength range are required in order to understand the UVO spectra of 
XTINSs and their connection to the X-ray spectra.

The only existing HST observation of 
an XTINS in a near-IR (NIR) band brought another surprise. 
An HST observation of RX\,J0806.4--4123 (J0806 hereafter) in the $\lambda = 1.4$--1.7\,$\mu$m 
not only showed a factor of 500 (!) excess
over the X-ray BB extrapolation,
but also 
resolved its {\em extended component}, with a size of about $1''$ and 
a $\gtrsim 50\%$ contribution to the total 
flux in this band 
\citep{Posselt2018}.
These authors concluded that the extended NIR emission could be the first NIR-only 
pulsar wind nebula (PWN)
or the first resolved fallback disk  around an isolated NS. It is, of course, very interesting to observe other XTINS at (N)IR wavelengths to see how 
common this phenomenon is. 

\subsection{Previous observations of RX J2143.0+0654}
\label{sec:previous_observations}
Of the 5 XTINSs observed by 
\citetalias{Kaplan2011} in 2 HST bands (HST program \#11654), particularly strong discrepancy between the observed UVO spectral slope and the R-J slope  was found for RX\,J2143.0+0654 (J2143 hereafter), also known as RBS\,1774.
Therefore, we chose this object for a more detailed investigation of an XTINS UV-optical-IR (UVOIR) spectrum with the goal to understand the origin of the UVOIR emission from these puzzling objects.

J2143 was discovered with the ROSAT X-ray observatory \citep{Zampieri2001}.
Observations 
with the European Photon Imaging Camera (EPIC) aboard the XMM-Newton X-ray observatory allowed 
\citet{Zane2005} to detect pulsations with a period $P=9.437$ s and a pulsed fraction
with semiamplitude of $3.6\pm0.6$\%.
Fitting the spectrum with a BB continuum plus a Gaussian absorption line, these authors estimated 
a BB temperature 
$kT=104.0\pm 0.4$ eV and an unabsorbed flux $F_{\rm 0.2-2\,keV}^{\rm unabs} \approx 5.2\times 10^{-12}$ erg cm$^{-2}$ s$^{-1}$. 
 They also  
 estimated the magnetic field, $B\sim 1.4\times 10^{14}$ G, assuming that the absorption feature at about 0.7 keV is due to proton cyclotron absorption. 
  Using data from the Reflection Grating Spectrometer (RGS), \citet{Cropper2007} suggested an additional absorption feature at 0.4 keV in the J2143's spectrum.
 
\citet{Kaplan2009b} analyzed a larger set of XMM-Newton EPIC data and found spectral properties similar to those reported by \citet{Zane2005} and \citet{Cropper2007}, including possible spectral features at 0.75 keV and 0.4 keV. These authors estimated the period derivative, consistent with $B\sim 2\times 10^{13}$ G, much lower than estimated by \citet{Zane2005}.
 
 \citet{Schwope2009} found another 
 spectral model, two BBs with $kT_{\rm hot}\approx 104$ eV and $kT_{\rm cold}\approx 40$ eV, which marginally fits the EPIC + RGS data. Finally, from the analysis of several EPIC observations plus deep (210 ks) SRG/eROSITA observations, \citet{Pires2023} suggested a model that consists of two BBs with $kT_{\rm hot}=107\pm3$ eV, 
 $kT_{\rm cold}\approx 43$ eV, 
 and three spectral features centered at 0.39, 0.55 and 0.74 keV. These authors also reported a
 decrease in pulsed fraction 
 to 2.5\%. 
 
 \citet{Bogdanov2024} presented a thorough timing analysis of XMM-Newton, Chandra, and NICER observations of J2143 in 2004--2023. Their timing solution gives a complicated pulse shape with 3 peaks per period, whose positions depend on photon energy. The measured period derivative $\dot{P}= (4.145\pm0.005) \times 10^{-14}$ s s$^{-1}$ 
 corresponds to $\Edot=2.0\times 10^{30}$ erg s$^{-1}$, $B\sim 2.0\times 10^{13}$ G, and $\tau_c \sim 3.6$ Myr.

\begin{deluxetable*}{lllccc}[ht]
\tablecaption{HST observations of RX\,J2143.0+0654}
\tablehead{
\colhead{Start time (UT)} & \colhead{Instrument} &\colhead{Filter}  & \colhead{$\lambda_{\rm piv}$} & \colhead{$W_{\rm eff}$}  & \colhead{Exposure} \\
\colhead{} & \colhead{} & \colhead{} & \colhead{\AA} & \colhead{\AA}  & \colhead{s}}
\startdata
2024-05-16 19:40:28 & WFC3/UVIS2  & F300X & 2813 & 747  & 4926\\
2024-05-17 06:43:53 & ACS/SBC  & F140LP & 1519 & 259  & 2610 \\
2024-05-18 20:39:42 & ACS/WFC & F475W  & 4747 & 1272  & 4184 \\
2024-06-18 16:14:02 & ACS/SBC & F140LP & 1519  & 259   & 2488 \\
2024-06-30 19:05:54 & WFC3/IR & F160W & 15369 & 2750  & 2497\\
2024-06-30 19:18:04 & WFC3/IR & F110W & 11534 & 3857 & 2347\\
\hline
2010-05-19 19:33:09 & ACS/WFC & F475W & 4747 & 1272  & 7076 \\
2010-05-23 14:38:16 & ACS/SBC & F140LP &1519 &259  & 8296 \\
\enddata
\tablecomments{
For the WFC3/IR observations, two F110W exposures were placed between two F160W exposures in each of the two orbits (see Section~\ref{sec:hstobs}). 
The start times for these observations are for the first (of four) exposures in each filter, while the exposure times are for the summed F160W and F110W exposures.
The last two rows
refer to the observations reported by 
\citetalias{Kaplan2011}.The wavelength $\lambda_{\rm piv}$ is the pivot wavelength as provided by HST,
the effective widths (defined as  in \citealt{Rodrigo2012})
 are taken from 
 \url{http://svo2.cab.inta-csic.es/svo/theory/fps/index.php?mode=browse}.}. 
\label{tab:HSTobservations}
\end{deluxetable*}
 
 Faint optical emission from this NS  ($m_B\sim 27$) was detected 
with the Very Large Telescope (VLT) 
\citep{Zane2008}
and Large Binocular Telescope (LBT)
\citep{Schwope2009}, with 
photometry results discrepant by half a magnitude, which might suggest an optical variability.
\citet{Posselt2009}
found a limit $H>22.0$ mag ($f_\nu < 1.55$ $\mu$Jy at $1.6$ $\mu$m) in a VLT observation.

The most precise measurements in the UVO range have been obtained with the HST by \citetalias{Kaplan2011}:
$f_\nu = 80\pm 5$ nJy at 
4709 \AA\ 
(ACS/WFC F475W filter) and $f_\nu = 
94\pm 7$ nJy at 1517 \AA\ (ACS/SBC F140LP filter), the latter value is corrected with account for the recent update on  SBC sensitivity 
\citep{Avila2019}.

Exceeding the X-ray  
BB extrapolation by at least a factor of 50, J2143 has the highest optical excess among all known XTINSs.
For a plausible optical extinction 
$A_V=0.12$,
the de-reddened flux densities are
$f_\nu = 95\pm 6$ nJy and $124\pm8$ nJy at pivot wavelengths 4747 \AA\ and 1519 \AA,
respectively. They  can be connected by a power-law (PL)  $f_\nu \propto \nu^\alpha$, $\alpha = 0.23\pm 0.08$. 

A source spectrum of an unknown shape 
cannot be inferred from only 2 observational points.
Therefore, we carried out
new HST observations that cover the 
UVOIR range with five broad filters (HST program \#17476; 
\citealt{Pavlov2023}). 
To investigate the connection of the UVOIR spectrum with the X-ray spectrum, we also requested a contemporaneous 
Target of Opportunity (ToO) observation 
with the XMM-Newton observatory, 
which was 
needed to 
evaluate
effects of possible long-term X-ray variability on the multiwavelength properties of the target.
Results of these observations are presented below.

\section{Observations and data reduction}
\subsection{HST observations}
\label{sec:hstobs}
\begin{figure*}[ht]
\begin{center}
\includegraphics[width=16cm]{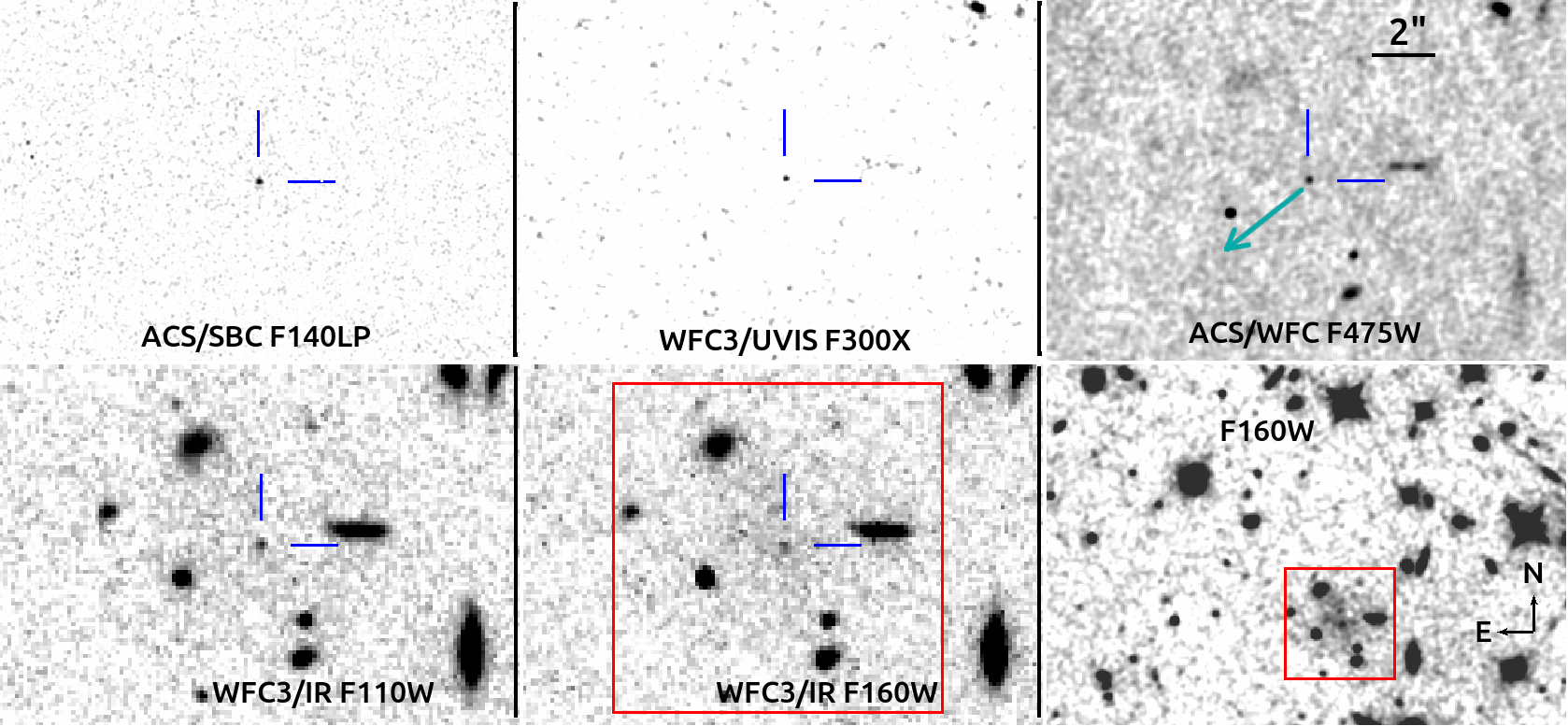}
\end{center}
    \caption{Images of the J2143 field, observed in five HST filters; north is up, east to the left. 
    The position of the pulsar is indicated by blue marks.
    Except for 
    the bottom right, the images are of the same angular sizes, $\approx16''\times 11''$. To illustrate the enhanced  emission near the location of the pulsar, the bottom right image shows a zoom-out of the WFC3/IR F160W image 
    in the middle-bottom panel, with the red box marking the same sky area.  
    The F475W (top right) image also shows the (dark-cyan) proper motion vector (see Section \ref{sec:propmot}).}
    \label{fig:HST_images}
\end{figure*}

We observed J2143 in 2024 May 16 -- June 30 in 5 spectral bands 
(8 HST orbits; see Table \ref{tab:HSTobservations}\footnote{See \url{https://www.stsci.edu/hst/phase2-public/17476.pdf} for technical details.
}). To check for variability and measure the proper motion, we  repeated the observations in the far-UV (FUV) F140LP 
and optical F475W 
filters (2 orbits for each of the filters).
In addition, we observed the target in a very broad F300X near-UV (NUV) filter (2 orbits) and in the F110W and F160W NIR filters. 
 Since the F110W filter can suffer from unpredictable background variations from the He\,I airglow in the upper atmosphere\footnote{ See WFC3 Instrument Science Report 2014-03 \url{https://www.stsci.edu/files/live/sites/www/files/home/hst/instrumentation/wfc3/documentation/instrument-science-reports-isrs/_documents/2014/WFC3-2014-03.pdf}.}, we used a  ``sandwich observing strategy'' for mitigation,
placing the F110W exposures 
in the middle, and the F160W exposures at the beginning and end of each of the two orbits.
Images of the target vicinity in the 5 filters are shown in Figure \ref{fig:HST_images}.
We 
used the drizzled images processed by the standard pipelines (calibration software caldp$\_$20240509, CALWF3  version 3.7.1 for the WFC3 data, and AstroDrizzle Version 3.7.0.). The pixel drizzle fractions are 1 for F110W, F160W, F140LP, and 0.8 for F300X and F475W; for details on the drizzling parameters see the DrizzlePAC software documentation \citep{Gonzaga2012,Fruchter2010}.

\subsection{XMM-Newton observations}
The XMM-Newton observations (ObsID 0935190301) were carried out on 2024
May 14 for 34.9\,ks. Thin filters were used for all EPIC detectors. 
The EPIC-pn detector was in Large Window mode (frame
time 47.7\,ms), MOS1 and MOS2 in Large Window mode (frame time 0.9\,s).  
We used XMM SAS version 21.0 for data analysis. 

\section{Data analysis}
\label{sec:data_analysis}

\subsection{UVOIR spectrum of the point source counterpart}
\label{sec:uvoir_spectrum}
\begin{deluxetable*}{lrccccccrr}[ht]
\tablecaption{Photometry of the HST observations of RX\,J2143.0+0654 \label{tab:hst_photometry}}
\tablehead{
\colhead{Filter} & \colhead{$t_{\rm exp}$} &\colhead{$r_{\rm extr}$}  & \colhead{$\phi$} &
\colhead{$N_{\rm tot}$} & 
\colhead{$N_{\rm bgd}$} & 
\colhead{$N_s$} & 
\colhead{$C_s$} & 
\colhead{${\cal P}_\nu$} & 
\colhead{$\langle f_\nu \rangle$} 
\\
\colhead{ } & \colhead{s} &\colhead{arcsec}  & \colhead{$\%$ } & \colhead{cnts} & \colhead{cnts}& \colhead{cnts} & cnt/s & \colhead{nJy\,s/cnts} & \colhead{nJy} 
}
\startdata
F140LP & 5098 & 0.175 & 60& 229.5 & $19.9 \pm 5.3$ & $210 \pm 15$ & 
$0.0685\pm 0.0050$ & 1651 & $113.1\pm 8.3$  \\
F140LP (K11) & 8296 & 0.175 & 60 & 368.3 & $29.9 \pm 5.6$ & $338 \pm 19$ & 
$0.0680\pm 0.0064$ &1593 & $108.5 \pm 6.2$ \\
F140LP (our+K11)\tablenotemark{a} & 13394 & 0.175 & 60 & 585.0 & $48.5\pm7.6$ & $537\pm24$ & 
$0.0668\pm 0.0030$ & 1651 & $110.2\pm5.0$ \\
F300X  & 4926 & 0.15 & 76 & 611.8 & $97 \pm 69$ & $515 \pm 72$ & 
$0.138\pm 0.019$ & 350 & $48.4 \pm 6.8$ \\
%
%
F475W  & 4184 & 0.2 & 84 & 11845 & $10166 \pm 210$ & $1678 \pm 214$ & 
$0.478\pm 0.061$ & 139 & $66.4 \pm 8.5$ \\
F475W (K11) & 7076 & 0.2 & 84 & 20744 & $17478 \pm 184$ & $3665 \pm 192$ & 
$0.617\pm0.032$ & 138 & $75.7 \pm 4.5$\\
F475W (our+K11) & 11260 & 0.2 & 84 & 32445 & $27524\pm279$ & $4922\pm 287$ & 
$0.520\pm0.030$ & 139 & $72.3\pm4.2$ \\
F110W & 2347 & 0.47 & 86 & 107758 & $104975 \pm 594$ & $2783 \pm 596$ & 
$1.38\pm 0.30$ & 68 & $93 \pm 20$ \\
F160W & 2497 & 0.47 & 85 & 61529 & $59082 \pm 482$ & $2446 \pm 485$ & 
$1.15\pm 0.23$ & 153 & $177 \pm 35$ \\
\enddata
\tablenotetext{a}{Numbers of counts in the K+11 observation are divided over 1.036 to compensate for a higher SBC sensitivity in that observation.} 
\tablecomments{
The rows marked `K11' provide results of our photometry of the 
\citetalias{Kaplan2011} data, the `our+K11' rows combine results from our and \citetalias{Kaplan2011} observations.
The column $\phi$ provides the fractions of source counts in the apertures with radii $r_{\rm extr}$,
$N_{\rm tot}$ is the total number of counts (electrons)  in the source aperture, 
$N_{\rm bgd}$
and $N_{s}$
are the background and net source count numbers in the chosen aperture,
$C_s$
is the aperture-corrected source count rate,
${\cal P}_\nu$ is the count rate-to-flux conversion factor,
and $\langle f_\nu\rangle$
is the 
characteristic flux density in the filter passband.
}
\end{deluxetable*}
\subsubsection{Photometry}
\label{sec:photometry}
Photometry of J2143 in 5 HST filters is presented in Table~\ref{tab:hst_photometry}, which
provides the radii $r_{\rm extr}$ of the source apertures chosen and the corresponding fractions $\phi$ of point source counts in these apertures, 
taken from the respective online instrument handbook tables of encircled energy fractions. In each of the source apertures we measured the total (source + background) count rate $C_{\rm tot}$ (in units of electrons/s) and converted it to the number of counts (electrons) $N_{\rm tot}=C_{\rm tot} t_{\rm exp}$. 
To measure the background and estimate the source count 
uncertainty, we used the ``empty aperture'' approach 
(e.g., \citealt{Skelton2014}).
The background counts 
were measured in multiple apertures (from 40 to 115, depending on filter) 
of the same 
size as the source aperture, uniformly distributed in source-free background regions. 
We calculated the mean $\overline{N}_{\rm bgd}$ and the variance $\sigma_{N_{\rm bgd}}^2$  of 
measured numbers of counts 
in the background apertures, 
and obtained the net source count 
number $N_{s}$ and its uncertainty $\sigma_{N_s}$:
\begin{equation}
\label{eq:netfluxerr}
   N_s = N_{\rm tot} -\overline{N}_{\rm bgd},\quad\quad \sigma_{N_s} = \left(\sigma_{N_{\rm bgd}}^2 + N_s  
    \right)^{1/2}, 
\end{equation}
where 
$N_{\rm tot}$ in the total number of counts in the source aperture.

To convert the aperture-corrected pulsar's count number
measured in a given filter,
$N_{s}/(t_{\rm exp}\phi) = C_s$, 
to the 
mean 
flux density in that filter, 
\begin{equation}
   \langle f_\nu\rangle \equiv 
   \frac{\int f_\nu(\nu) {\cal T}(\nu) \nu^{-1}\,d\nu}{\int {\cal T}(\nu) \nu^{-1}\,d\nu} =
     C_s{\cal P}_{\nu}\,,
\label{eq:mean_flux}
\end{equation}
we used the inverse sensitivity
\begin{equation}
    {\cal P}_\nu = h \left[A_{\rm tel} \int {\cal T}(\nu) \nu^{-1}d\nu\right]^{-1},
    \label{eq:conversion_factor}
\end{equation}
where $h$ is the Planck constant, $A_{\rm tel}$ is the collecting area of the telescope, and ${\cal T}(\nu)$ is the dimensionless bandpass throughput function\footnote{See \url{https://www.stsci.edu/files/live/sites/www/files/home/hst/documentation/_documents/SynphotManual.pdf}, Chapter 7.}.
For an observation with a given filter, the conversion factor ${\cal P}_\nu$ is provided as the keyword {\tt photfnu} in the data file header. Note that the 
mean flux 
flux density defined by Equation (\ref{eq:mean_flux}) is fully determined by the measured count rate and filter properties,
and it is not tied to a specific wavelength within the passband.

To look for flux changes in the 14 years since the \citetalias{Kaplan2011} observations, we applied the above-described approach to (re)measure the 
F140LP and F475W flux densities 
in those data 
and put the results in Table \ref{tab:hst_photometry}.  Since we found no statistically significant changes in these filters, we 
combined 
the old and new observations to reduce the uncertainty. 
\subsubsection{Extinction and distance towards J2143}
\label{sec:extinction}
\begin{figure}[ht]
\includegraphics[width=8.5cm]{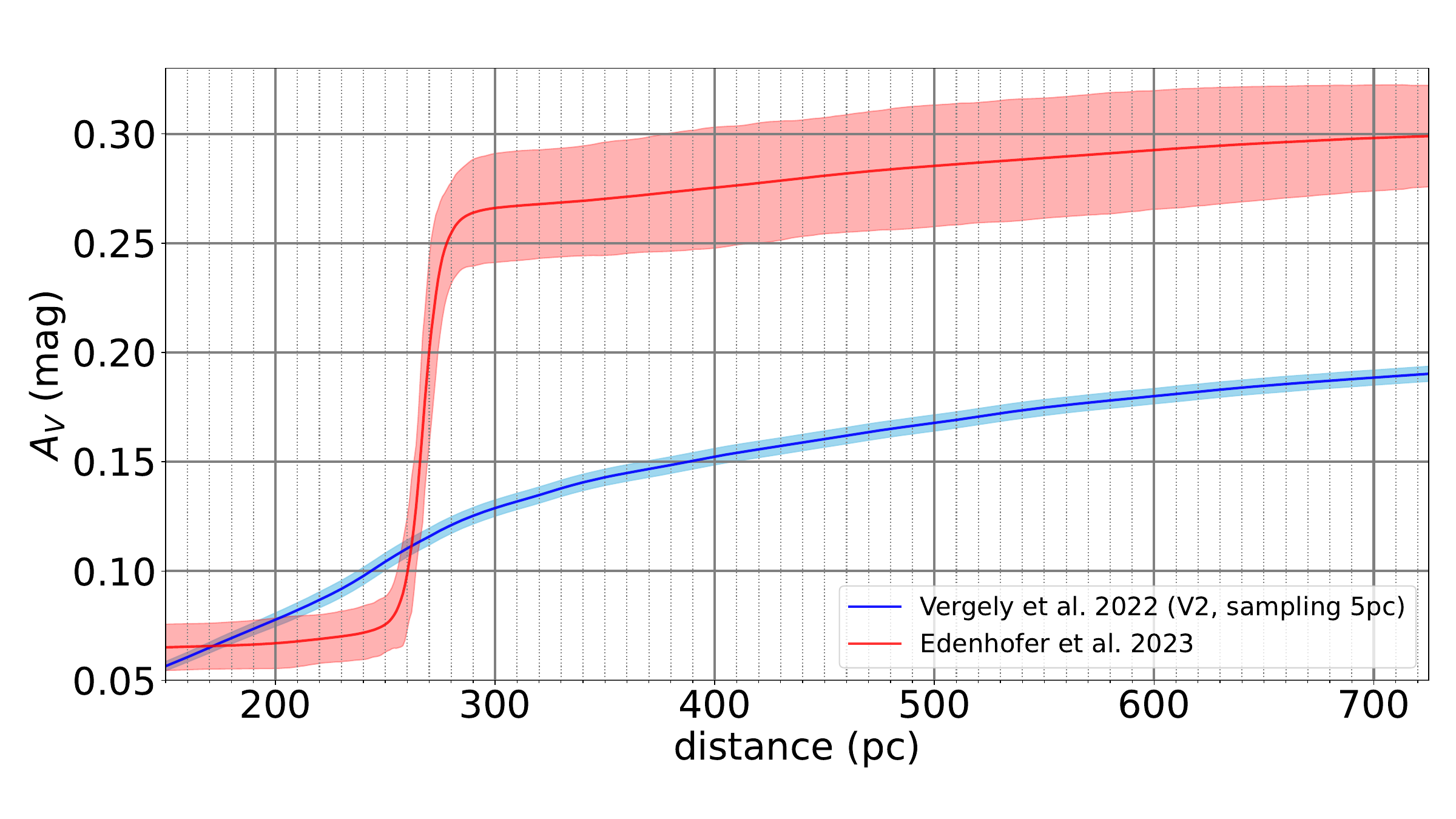}
\caption{Dependence of the extinction coefficient $A_V$ 
on distance in the direction to J2143. The blue line corresponds to data from \citet{Vergely2022}, 
obtained via the G-Tomo tool \citep{Lallement2022} that is available at ESA Datalabs. 
The red line shows the model by \citet{Edenhofer2024}. Shaded regions mark the $3\sigma$ uncertainty ranges.
}
\label{fig:extinction_distance}
\end{figure}
The shape of the observed UVOIR spectrum 
is affected by wavelength-dependent 
extinction, caused by interstellar dust.
With only 5 spectral bands and strong correlation of the extinction coefficient with spectral model parameters, we cannot measure the extinction from the UVOIR spectral fitting alone, but we may attempt to constrain it using other methods.

Extinction in a given direction depends on distance.
This dependence can be found from 3D extinction maps
\citep{Green2019,Leike2020,Lallement2022,Vergely2022,Edenhofer2024}.
Different maps, however, may yield different results in some directions because of different angular and distance resolution and different methods used to construct the maps.
For the direction toward J2143 (Galactic coordinates $l=62\fdg66$, $b=-33\fdg14$),
this is demonstrated in Figure~\ref{fig:extinction_distance}, which shows the distance dependencies of the cumulative extinction coefficient $A_V$ in the $V$ band. 
The $A_V(d)$ curves with $3\sigma$ uncertainties, extracted from the maps by \citet{Vergely2022} and \citet{Edenhofer2024}, 
are shown in blue and red, respectively. The main difference between 
them is the presence of a dust structure, centered at $d\sim 260$--270 pc, in \citet{Edenhofer2024}, which leads to a higher cumulative extinction at larger distances. A hint of a substantially larger
diffuse structure at similar distances is barely seen in the extinction dependence obtained from the \citet{Vergely2022} map, which has a lower resolution. Interestingly, the $A_V(d)$ curves obtained from the maps of \citet{Leike2020} and \citet{Green2019} are very close to \citet{Vergely2022} curve, but the high-resolution dust emission data obtained with the Planck observatory show a complex structure in that direction and $A_V\approx 0.27$ at infinity
(R.\ Lallement, priv.\ comm.). Given the higher resolution of the \citet{Edenhofer2024} map and its consistency with the Planck map, we consider this $A_V(d)$ dependence more reliable and use it in the following analysis.

The distance to J2143 has not been directly measured.
\citet{Zane2005} suggest $d\sim 280$ pc from the assumption that its X-ray luminosity is close to a ``typical'' XTINS luminosity,
$L_{\rm 0.1-2.4\,keV}\sim 5\times 10^{31}$ erg s$^{-1}$.
Given the large scatter of XTINS luminosities, this distance estimate can be off by a factor of $\sim2$--3.
However, extinction (and hence distance) can be constrained from the
hydrogen column density $N_{\rm H}$ estimated from X-ray spectral fits, 
using a linear correlation between $N_{\rm H}$ and $A_V$ 
(\citealt{Foight2016}, and references therein).
A problem with this approach is the 
strong dependence of $N_{\rm H}$ on 
model used for fitting the X-ray spectra of J2143 -- most of the explored models give\footnote{There is an outlier $N_{\rm H,20} =13.2\pm0.9$ in \citet{Pires2023}, which was likely caused by poorly known calibration of the eROSITA detectors at low X-ray energies. } $N_{\rm H,20} \equiv N_{\rm H}/(10^{20}\,{\rm cm}^{-2}) \approx2$--5. Using the correlation $N_{\rm H,20} = (28.7\pm 1.2) A_V$ \citep{Foight2016}, this range of $N_H$ corresponds to $A_V\approx0.07$--0.18. 
This broad extinction range corresponds to a relatively narrow distance range, $d\approx 240$--270 pc, as follows from the \citet{Edenhofer2024} curve $A_V(d)$  
in Figure \ref{fig:extinction_distance}. Thus, even with account for the large scatter in the $N_{\rm H}$-$A_V$ correlation, particularly strong at small distances, it looks plausible that J2143 is a NS embedded in the nearby dust cloud. In further analysis, we will investigate the dependencies of the J2143 parameters on extinction and distance around the fiducial values $A_V=0.12$ and $d=260$ pc.

\subsubsection{UVOIR spectral fits}
\label{sec:uvoir_fits}
\begin{figure}[ht]
\includegraphics[scale=0.21]{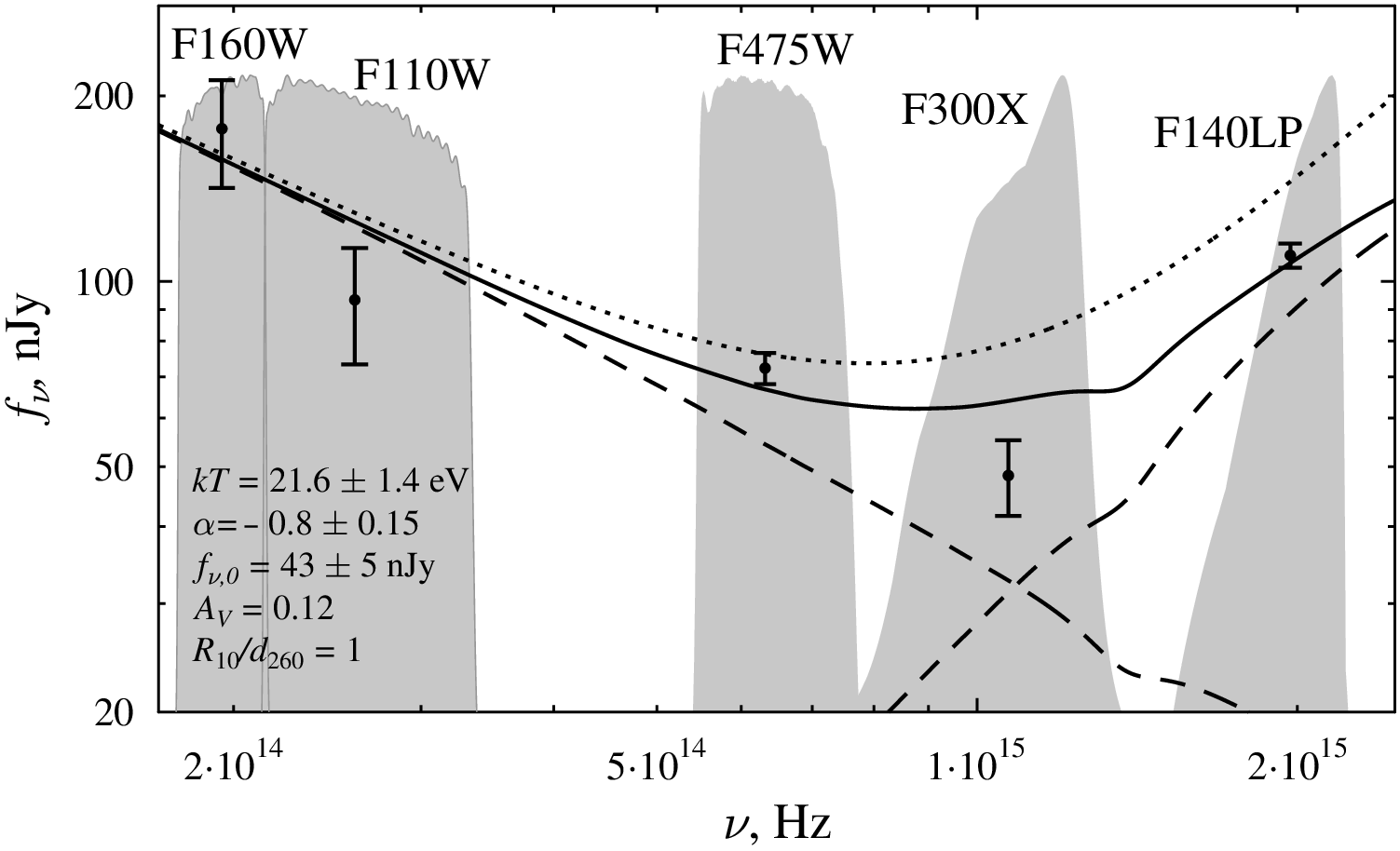}
    \caption{UVOIR spectrum of J2143. The data points with vertical error bars show the measured 
    mean 
    flux densities $\langle f_\nu\rangle$  with the statistical $1\sigma$ errors; they are plotted at 
    pivot wavelengths. The shaded areas show the filter throughputs (arbitrarily scaled). The solid curve shows the best fit for the PL+BB model (see Equation \ref{eq:PL+BB_model}) for $A_V=0.12$  at the fixed 
    $R_{10}/d_{260} = 1$. 
    The dashed curves show the PL and BB components. The dotted curve shows the dereddened best-fit spectrum. }
    \label{fig:uvoir_spectrum}
\end{figure}

\begin{figure}[ht]
\includegraphics[scale=0.19]{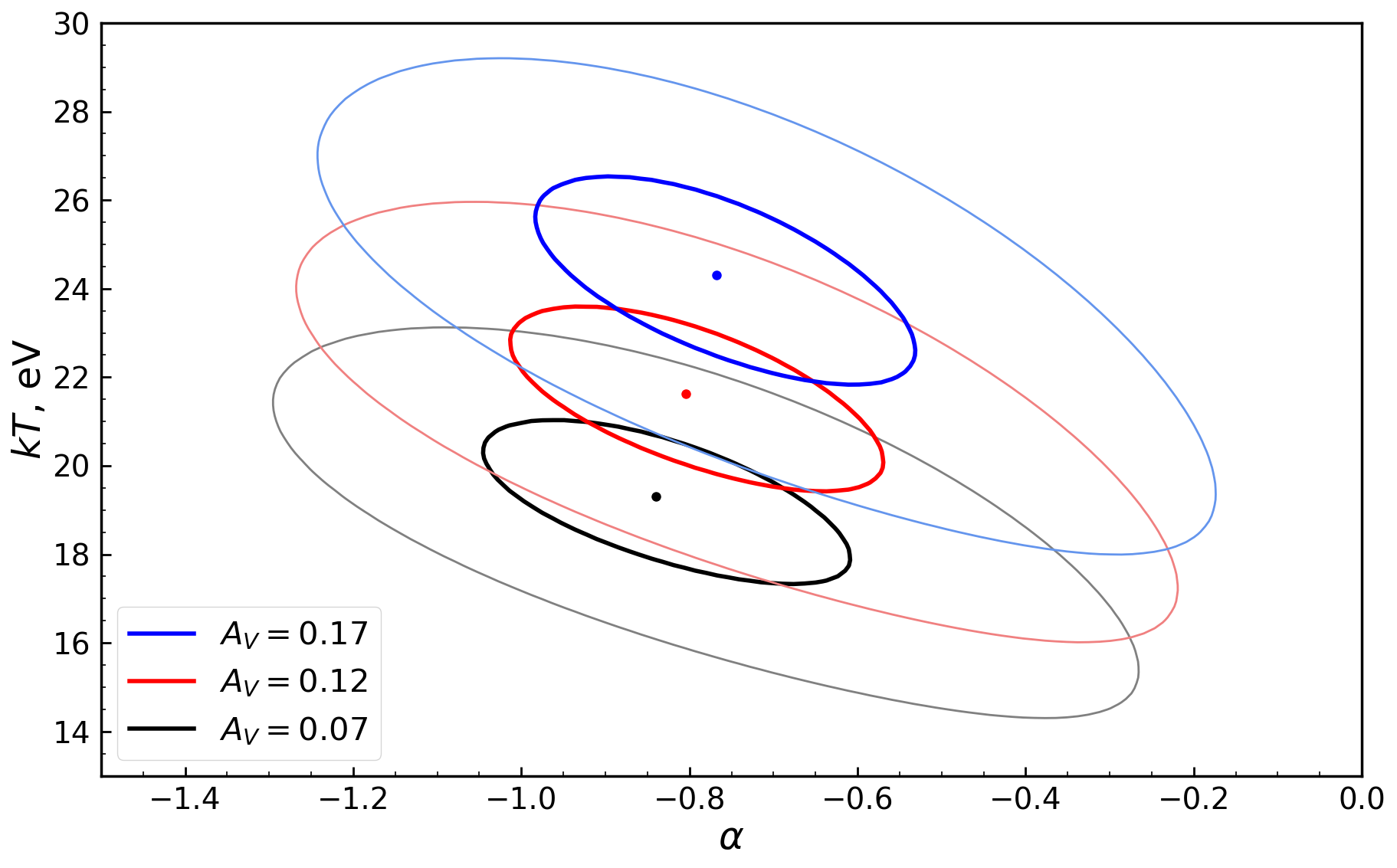}
\caption{Confidence contours for the UVOIR spectral fit with the PL+BB model (see Figure \ref{fig:uvoir_spectrum}) at 68.3\% and 99.7\% 
confidence levels (for two parameters of interest) in the $\alpha$-$kT$ plane for $R_{10}/d_{260} =1$ 
at the extinction values shown in the legend.}
\label{fig:conf_contours}
\end{figure}
%
To fit a spectral model $f_\nu^{\rm mod}$ to the measured count rates in the 5 filters, we 
calculate
model count rates as a function of model parameters,
\begin{equation}
    C_i^{\rm mod} = (A_{\rm tel}/h) \int f_\nu^{\rm mod} {\cal T}_i(\nu) \nu^{-1}\,d\nu\,,
    \label{eq:model_count_rate}
\end{equation}
with the aid of filter throughputs ${\cal T}_i(\nu)$ produced by {\tt synphot}, 
and find the parameter values that minimize the $\chi^2$ statistic
\begin{equation}
    \chi^2 = \sum_{i=1}^5 \frac{(C_i^{\rm mod} - C_i)^2}{\sigma_{C_i}^2}\,,
\end{equation}
where $C_i = N_{s,i}/(\phi_i t_{{\rm exp},i}) = C_{s,i}$ is the measured  source count rate in the $i$-th passband. 

As we immediately see from the data points in Figure 
\ref{fig:uvoir_spectrum}, the spectrum is qualitatively different from a simple PL or from a thermal spectrum.
It, however, can be described as a sum of a 
PL component and a R-J tail of a thermal component, at lower and higher frequencies, respectively, similar to UVOIR spectra of RPPs (see, e.g., 
\citealt{Abramkin2025},
and references therein).
Therefore, we will fit it with the absorbed PL + BB model,
\begin{equation}
    f_\nu^{\rm mod} = \left[f_{0} \left(\frac{\nu}{\nu_0}\right)^\alpha + 
    \frac{R_{\rm uv}^2}{d^2} \pi B_\nu(T_{\rm uv})\right] 10^{-0.4 A_\nu}\,,
\label{eq:PL+BB_model}
\end{equation}
where $\nu_0$, $f_0$ and $\alpha$ are the reference frequency, normalization, and slope of the 
PL component, $d$ is the distance, $B_\nu(T_{\rm uv})$ is the Planck function, 
and $R_{\rm uv}$ is the radius of 
an equivalent sphere in the UV range, i.e., a sphere with a uniform surface temperature $T_{\rm uv}$ that emits pure BB radiation. 
The surface temperature of a real NS can be nonuniform because of anisotropy of thermal conduction in a strong magnetic field, and its thermal emission can differ from the pure BB emission (e.g., \citealt{Perez-Azorin2005}). Therefore, Equation (\ref{eq:PL+BB_model}) should be considered as a phenomenological model in which $T_{\rm uv}$ is an estimate of an average surface temperature measurable in the UV range, while $R_{\rm uv}$ can be substantially 
smaller than the true NS radius (see Section 4.2 for further discussion). 

We choose $\nu_0=1\times 10^{15}$ Hz (corresponding to $\lambda_0=3000$ \AA). We take the frequency dependence of the extinction $A_\nu$ from 
 \citet{Gordon2023} 
and normalize it by the extinction value $A_V$ in the $V$ band, adopting $A_V=0.12$ as a fiducial value (see Section \ref{sec:extinction}).  

The thermal term in Equation (\ref{eq:PL+BB_model}) can be written as 
\begin{equation}
    \left(\frac{R_{\rm uv}}{d}\right)^2 \pi B_\nu(T_{\rm uv}) = 
7.2 \frac{(R_{10}/d_{260})^2 \nu_{15}^3}{\exp(h\nu/kT_{\rm uv}) -1}\,{\rm nJy}\,
    \label{eq:BB_term}
\end{equation}
where $R_{10} = R_{\rm uv}/10\,{\rm km}$, $d_{260}=d/260\,{\rm pc}$, and $\nu_{15}=\nu/10^{15}\,{\rm Hz}$. The model (\ref{eq:PL+BB_model})
has 5 parameters.
However, in the R-J regime, $h\nu\ll kT_{\rm uv}$, the 
model spectrum and the $\chi^2$ value depend on the so-called R-J parameter $kT_{\rm uv}(R_{\rm uv}/d)^2$ \citep{Pavlov1997} rather than on $T_{\rm uv}$ and $R_{\rm uv}/d$ separately. 
Since the R-J parameter strongly correlates with the extinction, we fit the model to the 5 data points at fixed extinction values, varying 3 model parameters.  

Figure \ref{fig:uvoir_spectrum} shows an example of the PL+BB fit at 
$A_V=0.12$, $R_{10}/d_{260} =1$. 
The best-fit PL component, $f_\nu = f_{0}\nu_{15}^{\alpha}$,  decreases with frequency -- e.g., 
$\alpha=-0.80 \pm 0.15$ at $A_V=0.12$. 
The PL slope slightly steepens with increasing extinction, as seen in Figure \ref{fig:conf_contours}, but for a reasonable extinction interval, $0.05 \lesssim A_V \lesssim 0.2$, the change of the best-fit $\alpha$ is smaller than its $1\sigma$ uncertainty.

The thermal component dominates in the UV range, at $\lambda \lesssim 3000$ \AA. 
Taking into account that the brightness temperature in the R-J regime is proportional to $(d/R)^2$, the fit shown in Figure \ref{fig:uvoir_spectrum} corresponds to 
$kT_{\rm uv} \sim 22 (d_{260}/R_{10})^2$ eV,
i.e., this brightness temperature is rather uncertain because 
the equivalent sphere radius is unknown. An additional (smaller) temperature uncertainty is due to extinction uncertainty.
According to Figure \ref{fig:conf_contours},  
the temperature slightly increases with increasing $A_V$. 
Additional constraints on the surface temperature and NS thermal emission
can be obtained from 
analysis of the 
X-ray emission and from joint fits of the UVOIR and X-ray spectra, provided below.

\subsection{Possible extended emission in the NIR bands}
\label{sec:extended}
As can be seen in Figure~\ref{fig:HST_images}, there seems to be a hint of flux enhancement around the pulsar in the NIR images. Using a polygon aperture (area of $\approx 13\,{\rm arcsec}^2$; 
see  Appendix~\ref{extmeas} for details), we measure  
net flux densities $f_{\rm F160W}^{\rm ext} = 
1.69\pm0.25$\,$\mu$Jy and 
$f_{\rm F110W}^{\rm ext} = 
0.88\pm0.17$\,$\mu$Jy in the F160W and F110W filters, respectively,
where the errors only reflect statistical uncertainties.
Neglecting possible contribution of unrecognized faint sources and assuming a 
PL spectral shape ($f_{\nu}^{\rm ext}\propto \nu^{\alpha_{\rm ext}}$), the 
supposed extended emission  has a spectral index  of $\alpha_{\rm ext} = -2.27 \pm 0.83$.

\subsection{XMM-Newton data analysis}
\label{sec:xmm_data_analysis}
\subsubsection{Timing analysis}
In order to check for possible variability of the X-ray pulsations of J2143, such as reported by \citet{Pires2023},
we carried out a 
timing analysis of the EPIC data.
We removed data obtained during the largest background flare and used the remaining
counts from the energy range 0.15--2\,keV (single and double patterns 
for pn and patterns 0--12 for MOS), 
extracted from circles with
$47''$ and $40''$ radii for the pn and MOS detectors, respectively.

We applied 
the binning-free Fourier analysis 
to barycenter-corrected times of arrival 
as
described in Appendices of \citet{Hare2021} and \citet{Posselt2024}.
This analysis starts from calculating 
the empirical Fourier coefficients, 
\begin{eqnarray}
    a_k = \frac{2}{N} \sum_{i=1}^N \cos 2\pi k\phi_i = s_k \cos 2\pi\psi_k\,, \\
    b_k = \frac{2}{N} \sum_{i=1}^N \sin 2\pi k\phi_i = s_k \sin 2\pi\psi_k\,,
\end{eqnarray}
where 
$s_k$ and $\psi_k$ are the amplitude and phase of $k$-th harmonic, $\phi_i = 
\nu_{\rm trial} (t_i-t_{\rm ref})$, $t_i$ is the time of arrival of the $i$-th count, $t_{\rm ref}$ is the reference epoch, and $\nu_{\rm trial}$ is the trial frequency. 
The empirical signal frequency $\nu$ is the trial frequency value 
at which the Fourier power $Z_K^2 = (N/2)\sum_{k=1}^K s_k^2$, summed over statistically significant harmonics, is maximal. 

In our case, 
 harmonics $k\geq 4$ are not statistically significant (i.e., $K=3$), and the empirical signal frequency is $\nu=0.1060619(13)$ Hz, fully consistent with the ephemeris derived by \citet{Bogdanov2024} from observations of 2004--2023. The frequency uncertainty, $\sigma_\nu=1.3\times 10^{-6}$ Hz, is estimated from Equation (A24) of \citet{Posselt2024}.

\begin{deluxetable}{lcccc}
\tablecaption{Fourier amplitudes and phases, and $Z_k^2$ statistics for the first 4 harmonics at $\nu=0.1060619$} \label{tab:xmm-timing}
\tablehead{
\colhead{$k$} & \colhead{1} & \colhead{2} & \colhead{3} & \colhead{4} 
}
\startdata
      &     &  pn + MOS1\&2 &   &   \\
\hline      
$s_k$ & 0.0269(49) & 0.0067(49)  & 0.0188(49)  & 0.0080(49) \\
$\psi_k$ & 0.076(29) & 
0.37(12) & 0.230(42) & 
0.58(10)\\
$Z_k^2$ & 29.5 & 31.4 & 45.7 & 48.2 \\ 
\hline
      &     &  pn only &   &   \\
\hline      
 $s_k$ & 0.0258(55) & 0.0099(55) & 0.0135(56) & 0.0071(56) \\
$\psi_k$ & 0.060(34) & 0.46(9) & 0.219(65) & 0.56(12) \\
$Z_k^2$ & 21.6 & 24.8 & 30.7 & 32.3 \\ 
\enddata
\tablecomments{The numbers in parentheses are $1\sigma$ uncertainties of the last digits of the measured quantities. 
The amplitude and phase uncertainties 
are estimated as $\sigma_{s_k} = (2/N)^{1/2}$
and 
$\sigma_{\psi_k} = (2\pi s_k)^{-1}(2/N)^{-1/2}$, where $N$ is the number of counts \citep{Posselt2024}.
}
\end{deluxetable}


Table \ref{tab:xmm-timing} shows the values of the empirical Fourier amplitudes and phases, as well as the sums of Fourier powers, 
for the first 4 harmonics, at the measured frequency. 
 In the upper half of the table the Fourier parameters were calculated from the pn + MOS1\&2 data ($N=81654$ counts, $T_{\rm span} = 33.2$ ks), which we used for measuring the frequency. Since the 
time resolution of the MOS data is too crude to analyze harmonics $k\geq 3$, in the lower half of the table we show the same Fourier parameters for the pn data only ($N=64848$ counts, $T_{\rm span}=31.3$ ks). Notice that $s_3>s_2$ in both cases, i.e., the third harmonic is more powerful than the second one.

\begin{figure}[ht]
\includegraphics[scale=0.23]{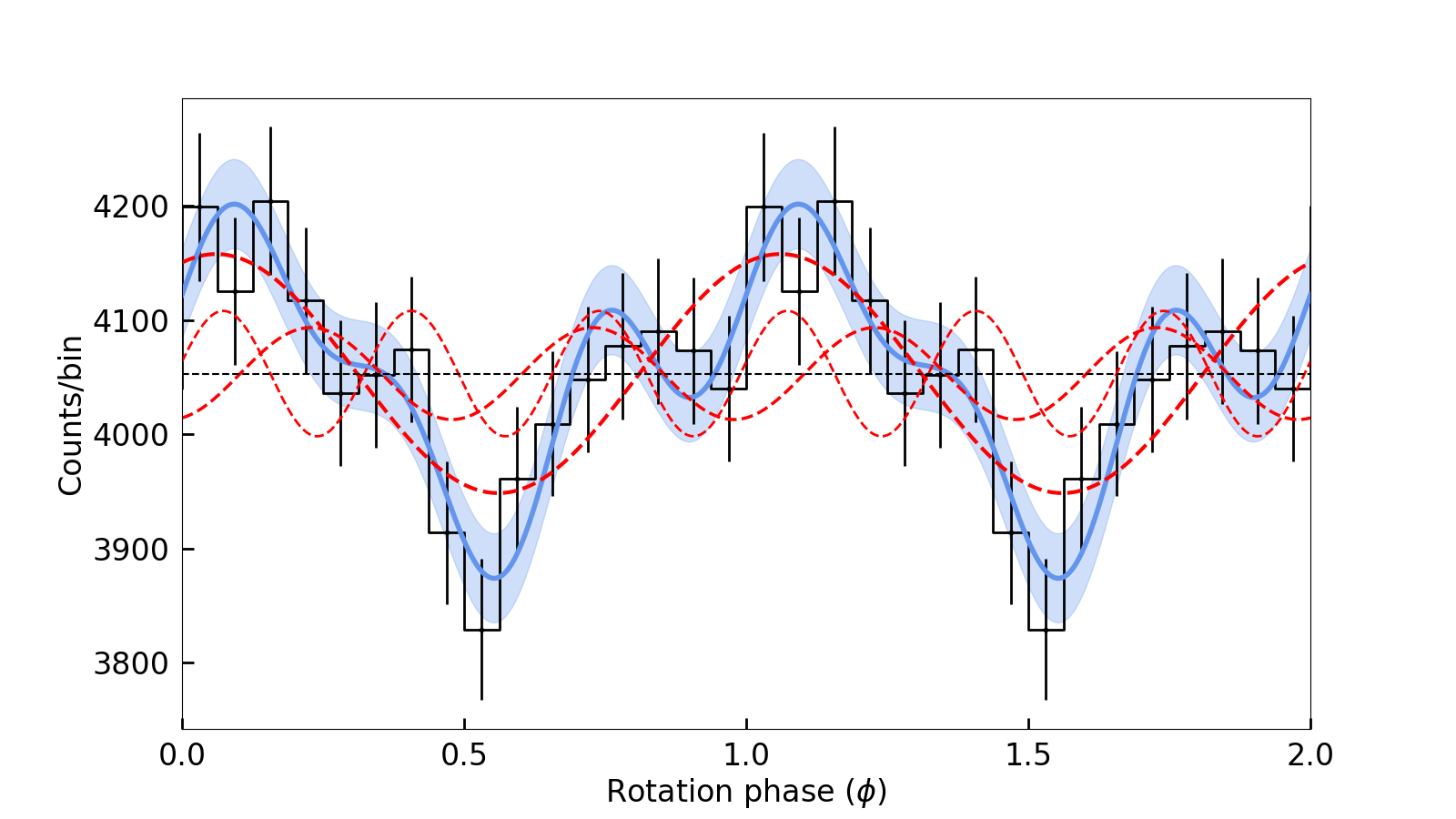}
\caption{
Pulsations of J2143 with frequency $\nu=0.1060619$ Hz in the 0.15--2 keV energy range,
 obtained from the EPIC pn 
data. 
The red dashed lines 
display the  $k\leq 3$ 
Fourier harmonics. The solid blue line shows their sum;
its 68\% uncertainty is indicated by the shaded area. 
The horizontal dotted line at  4053 
counts/bin  shows the constant component of the phase-folded light curve.
The zero phase corresponds to the epoch MJD 60443.6296296 (TDB).
The histogram demonstrates the 
more traditional binned profile, with 16 bins. 
}
\label{fig:pulse_profile}
\end{figure}

We used the  pn-only Fourier amplitudes and phases to plot the folded light curve (pulse profile) of J2143 (solid blue curve in Figure \ref{fig:pulse_profile}). 
The profile is very similar to that found by \citet{Bogdanov2024} from multiple observations, including 
possible smaller peaks at $\Delta\phi \approx \pm0.3$
with respect to the main peak. 

An example of a traditional binned light curve, plotted in the same figure, shows good agreement with the binning-free profile.

The Fourier amplitudes directly provide the rms pulsed fraction, $p_{\rm rms} = \left[\sum_k (s_k^2/2)\right]^{1/2}$ \citep{Hare2021}. 
Including the 
$k\leq 3$ terms in the sum, we obtain $p_{\rm rms} = 2.2(4)\%$.
%
Using the unbinned and binned light curves, we also calculated the area and amplitude pulsed fractions\footnote{
{ The area pulsed fraction, $p_{\rm area}$, is defined as the ratio of areas under the varying part of the light curve to the total area, while the amplitude pulsed fraction, $p_{\rm amp}$ (sometimes called “peak-to-peak” or “max-to-min” pulsed
fraction), is the difference between the maximum and minimum values of the pulse profile to their sum --- see \citet{Hare2021} for details.}}: 
$p_{\rm area}=4.4(9)\%$ and
$p_{\rm amp} = 4.1(8)\%$,
where the uncertainties 
were estimated from Monte-Carlo simulations. 
%
While the $3\sigma$ uncertainty range allows the lower $p_{\rm amp}=2.5$\% reported by \citet{Pires2023} from eROSITA observations, our results are closer to the previously reported $p_{\rm amp}\approx 4\%$ that were also based on XMM-Newton observations \citep{Zane2005,Kaplan2009b}. Thus, we do not see significant changes in the X-ray 
pulsations
of J2143 in the XMM-Newton observations.
%

\subsubsection{Spectral analysis}
\label{sec:xmm_spectral_analysis}
In order to 
 understand the multiwavelength emission from J2143,
we analyzed the spectra obtained by the EPIC detectors in our contemporaneous XMM-Newton observation, both separately and together with the UVOIR spectra.
After removing two background flares, the effective exposure times are 27.3\,ks and 23.5\,ks for the EPIC-pn and MOS1$\&$2 detectors, respectively. 
Pattern analysis with the task \texttt{epatplot} showed only very minor pileup in the target for EPIC-pn. To minimize 
effects of calibration uncertainties, we use only single events for EPIC-pn.   After employing the task \texttt{eregionanalyse}, the EPIC-pn and MOS1$\&$2 source spectra were extracted 
from circular regions with radii $47''$ and $40''$, and patterns 0 and 0--12, respectively. We chose background regions 
on the same 
chips as the source regions. 
Within an energy range of 0.2--2\,keV, we obtain net (source) count rates of 
$1.767 \pm 0.008$\,cps, 
$0.343 \pm 0.004$\,cps, and $0.313 \pm 0.004$\,cps for EPIC-pn, MOS1, and MOS2, respectively.  

We employ the XSPEC packgage (ver.\ 12.13.0) to fit the observed spectra. To describe the interstellar absorption, we use the {\tt tbabs} model with abundances from \citet{Wilms2000}. As no source counts are seen above 2 keV, we fit the spectrum in the 0.2--2 keV band. 
To compare with 
the previous works, we fit the spectra
with one- or two-component  
BB models ({\tt bbodyrad} in XSPEC) with up to two multiplicative {\tt gabs} models included to describe possible absorption features. 

%
\startlongtable
\begin{deluxetable*}{lccccccc}
    \tablecaption{Spectral fits of the 
    X-ray and X-ray + UVOIR data \label{tab:xmm_spectral_vert}}
\tablehead{
\colhead{ } & \colhead{1BB0G} & \colhead{1BB1G} & \colhead{1BB2G} & 
\colhead{2BB1G} & \colhead{2BB2G} & \colhead{2BB1G+PL} & \colhead{2BB2G+PL}
}
\startdata
$N_{\rm H}$ ($10^{20}$\,cm$^{-2}$) & $2.0 \pm 0.2$ & $2.1\pm 0.2$ & $1.9\pm0.2$ &  $9.0^{+2.5}_{-1.5}$ & $8.5^{+2.1}_{-1.8}$ & $2.6 \pm 0.5$ & $3.5^{+0.5}_{-0.8}$  \\
$kT_{\rm hot}$ (eV) & $105.9 \pm 0.5$ & $106.5\pm 0.5$ & $106.1\pm0.5$ &   $100\pm 2$ & $101\pm2$ &  $105.9 \pm 1.3$ & $105.2\pm0.5$ \\
$R_{\rm hot}/d_{260}$ (km) & $1.52 \pm 0.03$ & $1.52\pm0.03$ & $1.53\pm0.03$ & $2.2\pm0.2$ & $2.1\pm0.3$ & $1.56^{+0.11}_{-0.12}$ & $1.62^{+0.03}_{-0.04}$ \\
$L_{\rm hot}/d_{260}^2$ ($10^{31}$\,erg\,s$^{-1})$ & 3.7 & 3.8 & 3.8 & 6.3 & 6.0 & 4.0 & 4.1 \\
$kT_{\rm cold}$ (eV) & \ldots & \ldots & \ldots &  $29 \pm 2$ & $33^{+4}_{-3}$ & $38^{+4}_{-6}$ & $48^{+3}_{-8}$ \\
$R_{\rm cold}/d_{260}$ (km) & \ldots & \ldots & \ldots & $160^{+150}_{-80}$ & $70^{+100}_{-40}$ & $6.4^{+0.6}_{-0.5}$ & $5.9\pm 0.4$ \\
$L_{\rm cold}/d_{260}^2$ ($10^{31}$\,erg\,s$^{-1}$) &  \ldots & \ldots & \ldots & 230 & 80 & 1.1 &  2.3 \\
$E_{c1}$ (eV) & \ldots & $740\pm8$ & $739\pm8$ &  $753\pm 7$ & $752\pm7$ & $741 \pm 7$ & $741\pm 7$ \\
$\sigma_1$ (eV) & \ldots & $19^{+14}_{-15}$ & $22^{+13}_{-17}$ &  $50^{+16}_{-12}$ & $43\pm18$ & $22^{+12}_{-18}$ & $23^{+13}_{-19}$ \\
$s_1$ (eV) & \ldots & $18^{+30}_{-3}$ & $19^{+15}_{-3}$ &  $35^{+11}_{-7}$ & $30 \pm 10$ & $19^{+23}_{-3}$ & $18^{+55}_{-2}$ \\
$E_{c2}$ (eV) & \ldots & \ldots & $400\pm 11$ &  \ldots & $392\pm 17$ & \ldots & $403\pm 10$ \\
$\sigma_2$ (eV) & \ldots & \ldots & $5^{+18}_{-4}$ & \ldots & $<25$ & \ldots & $10^{+20}_{-9}$ \\
$s_2$ (eV) & \ldots & \ldots & $9^{+?}_{-3}$ & \ldots & $16^{+?}_{-12}$ & \ldots & $10^{+?}_{-4}$\\
$A_V = 0.035 N_{\rm H,20}$ & \ldots & \ldots & \ldots & \ldots & \ldots & $0.09 \pm 0.05$ & $0.12\pm0.03$ \\
$\alpha$ & \ldots & \ldots & \ldots &  \ldots & \ldots & $-0.78 \pm 0.16$ & $-0.74\pm 0.15$ \\
$f_{\nu_0}$ (nJy) & \ldots & \ldots & \ldots &  \ldots & \ldots & $44 \pm 7$ & $46\pm 6$ \\
$L_{\rm 1-10\,eV}/d_{260}^2$ ($10^{28}$\,erg\,s$^{-1})$ & \ldots & \ldots & \ldots &  \ldots & \ldots & 0.8 & 0.8 \\
$C_{\rm mos1}$ & $1.113 \pm 0.015$ & $1.109 \pm 0.015$ & $1.113 \pm 0.015$ & $1.107 \pm 0.015$ & $1.108 \pm 0.015$ & $1.109 \pm 0.015$ & $1.110\pm 0.014$ \\
$C_{\rm mos2}$ & $1.152 \pm 0.016$ & $1.153 \pm 0.016$ & $1.156 \pm 0.016$ & $1.151 \pm 0.016$ & $1.152 \pm 0.016$ & $1.153 \pm 0.016$ & $1.154\pm 0.016$ \\
$\chi_\nu^2$ [dof] & 1.82 [103] & 1.27 [100] & 1.15 [97] & 1.15 [98] & 1.15 [95] & 1.32 [101] & 1.22 [98] \\
\enddata
\tablecomments{The abbreviation mBBnG denotes the model with $m$ additive blackbody components and $n$ multiplicative {\tt gabs} components describing absorption lines for the Gaussian absorption coefficient; in the 2BBnG+PL models a UVOIR PL component is added. Each BB component is characterized by the temperature $kT$, equivalent sphere radius $R$ and bolometric luminosity $L=4\pi\sigma R^2 T^4$ for an assumed distance $d=260$ pc; the radii are calculated from the {\tt bbodyrad} norm $K_{\rm BB}=R_{\rm km}^2/d_{\rm 10\,kpc}^2$. The $i$-th {\tt gabs} component is characterized by the central energy $E_{ci}$, Gaussian width $\sigma_i$, and ``depth'' $s_i$ (see Equation (\ref{eq:gabs})).
For the PL component, $f_\nu = f_0 (\nu/\nu_0)^\alpha 10^{-0.4 A_V}$, we assume the optical extinction to be tied to the hydrogen column density, $A_V=0.035 N_{\rm H,20}$ \citep{Foight2016}, 
choose $\nu_0=10^{15}$ Hz ($h\nu_0=4.14$ eV), and calculate the UVOIR luminosity $L_{\rm 1-10\, eV}$. The coefficients $C_{\rm mos1}$ and $C_{\rm mos2}$ are additional normalization factors for the MOS count rates relative to the pn normalization. The last row provides the minimum reduced $\chi^2$ values and the numbers $\nu$ of degrees of freedom [dof]. Uncertainties indicate the
68\% confidence levels. }
\end{deluxetable*}

The fit results, presented in Table \ref{tab:xmm_spectral_vert}, are close to those obtained
by \citet{Zane2005}, \citet{Schwope2009}, \citet{Kaplan2009b}, and \citet{Cropper2007} from previous observations. 
The single BB model without spectral features, 1BB0G, leaves large residuals around 0.7 keV. Including one {\tt gabs} component (1BB1G model), 
\begin{equation}
    {\tt gabs}(E) = \exp\left[-\frac{s}{\sqrt{2\pi}\sigma}\exp\left(-\frac{(E-E_c)^2}{2\sigma^2}\right)\right]\,,
    \label{eq:gabs}
    \end{equation}
we obtain a better fit, with about the same BB temperature $kT\approx 106$ eV and radius $R\approx 1.5d_{260}$ km. 
The absorption line is centered at $E_c=0.74$ keV; its rather uncertain Gaussian width,
$\sigma\sim 4$--33 eV, 
is smaller than the energy resolution of the EPIC detectors. 
The third {\tt gabs} parameter $s$, called ``depth'' in the XSPEC description, is proportional to 
the optical depth $\tau$  at the line center,   $s\equiv (2\pi)^{1/2} \sigma \tau$,
and it is close to the line's equivalent width at $\tau\lesssim 1$.  
The $s$-$\sigma$ confidence contours show that $s\sim 15$--20 eV ($\tau \sim 0.1$--1) for more realistic $\sigma\sim 8$--33 eV, but it may reach $\sim 48$ eV at very low $\sigma\sim 4$ eV. 


Including a second {\tt gabs} line (1BB2G model), suggested by \citet{Cropper2007}, slightly improves the fit without a significant effect 
on the properties of the thermal component and the 0.74 keV line. The second line is centered at $\sim 0.4$ keV, and its $\sigma$ and $s$ parameters are even smaller than those of the 0.74 keV line. 

Following \citet{Schwope2009}, who suggested the presence of a second BB component with a lower temperature, we also 
added such a component for models with one and two {\tt gabs} lines (2BB1G and 2BB2G models). However, fits with these models are very unstable and the fitting parameters are very uncertain (so that we had to fix some of the fitting parameters to estimate uncertainties of others). The best-fit radii of the second BB component are unrealistically large, particularly for the 2BB1G models, which means that we should not trust other fitting parameters (e.g., the elevated $N_H$ values). Obviously, such models have too many parameters for the narrow energy range and the number of counts available.

We found that for each of the explored models, the observed (absorbed) energy flux is about the same, $F_{\rm 0.2-2\,keV} \simeq 2.8\times 10^{-12}$ erg cm$^{-2}$ s$^{-1}$, within 
a 1.4\% relative uncertainty. This value is consistent with the previous X-ray flux measurements, as well as the spectral parameters. 
Thus, the spectral and flux properties of J2143 at our observing epoch are consistent with those of the previous 20 years of XMM-Newton observations.

\subsection{
Joint spectral fit of the HST and XMM-Newton data}
\label{sec:joint_fit}
\begin{figure}
    \centering
\includegraphics[width=1.0\linewidth]{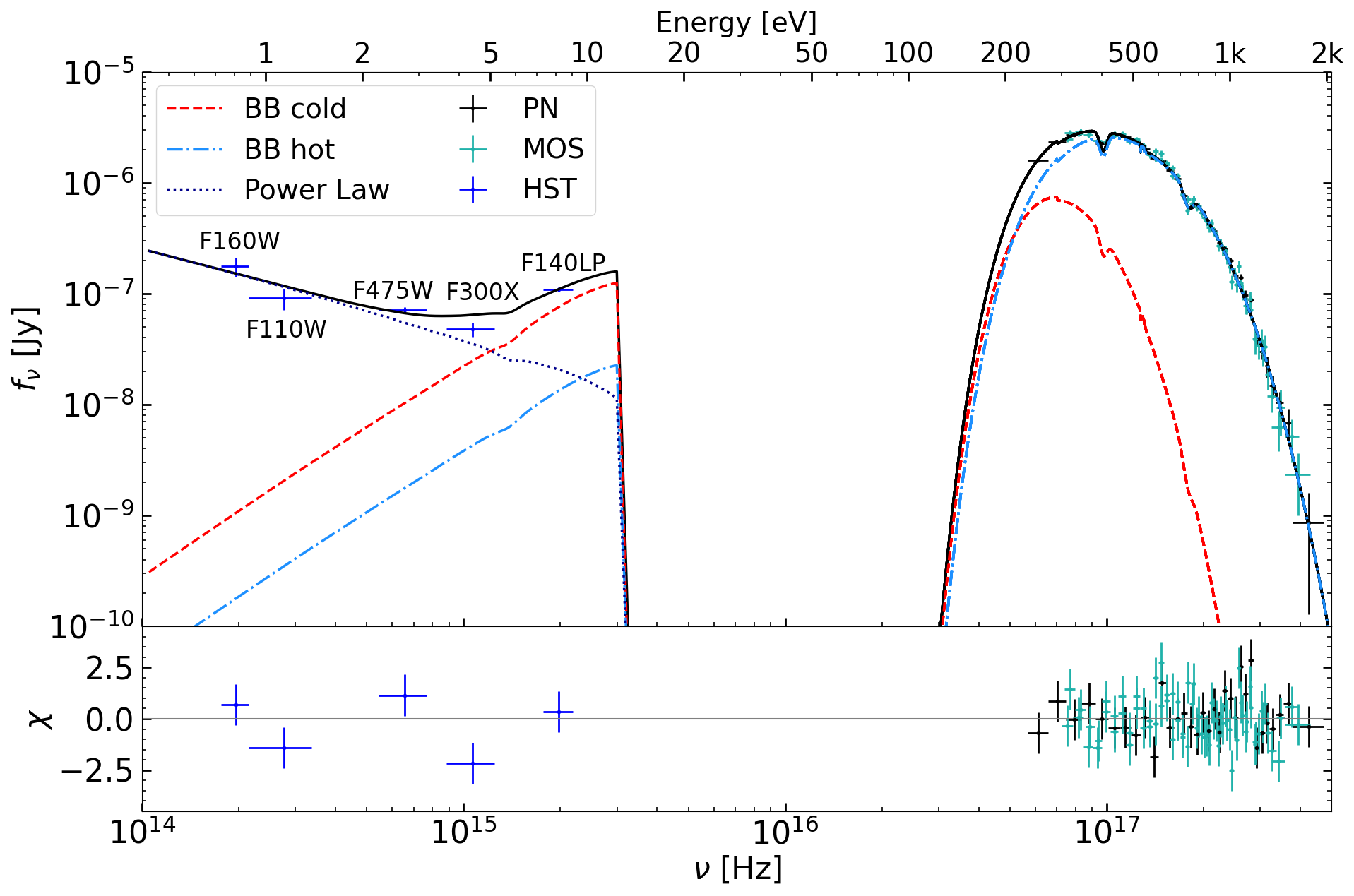}
    \caption{
    Unfolded XMM-Newton's EPIC and  HST spectrum of J2143 compared with the best-fit 
    2BB2G+PL model that includes the ISM absorption/extinction, 
    BB$_{\rm hot}$ and BB$_{\rm cold}$ components, two absorption features in soft X-rays, and the UVOIR PL component. The fit parameters are listed in Table 
    \ref{tab:xmm_spectral_vert}. The lower panel shows fit residuals defined as $\chi= ({\rm data}-{\rm model})/{\rm error}$.
    }
    \label{fig:mw_observed}
\end{figure}
\begin{figure}
    \centering
\includegraphics[width=1.05\linewidth]{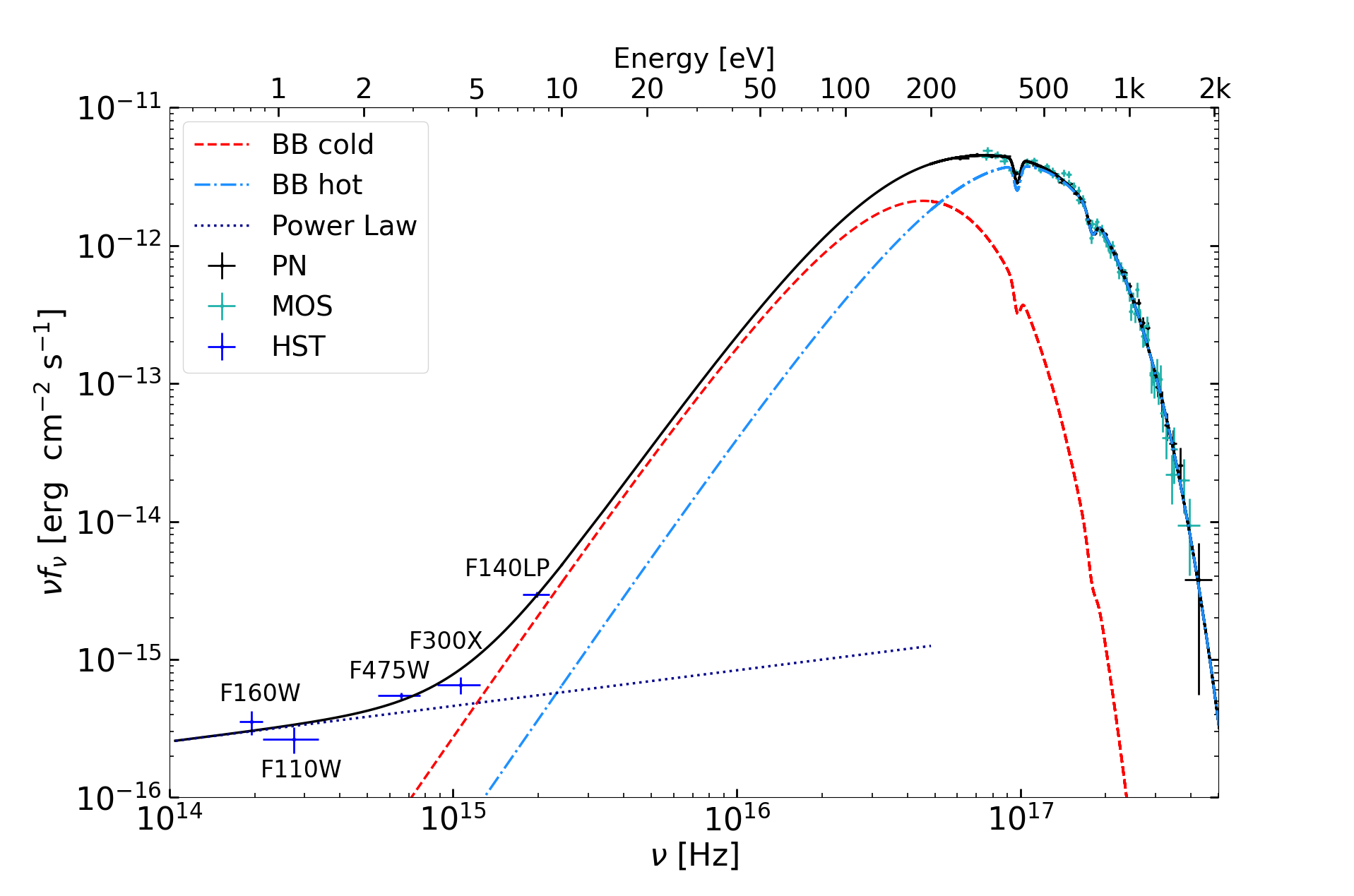}
    \caption{Spectral energy distribution (SED) 
    for the UVOIR through X-ray emission 
    for the same model as in Figure \ref{fig:mw_observed}.}
    \label{fig:mw_unabs}
\end{figure}
%
Assuming that the same two BB components
can contribute to the flux spectrum at both X-ray and UVOIR frequencies 
we
added a UVOIR PL component to the 
2BB1G and 2BB2G models and fitted the combined models in the 0.5\,eV -- 2\,keV range.
The fitting parameters for these models are shown in the two last columns of Table \ref{tab:xmm_spectral_vert},
while the UVOIR through X-ray absorbed energy flux spectrum $f_\nu$ and the spectral energy distribution (SED) $\nu f_\nu^{\rm unabs}$ are shown in  Figures \ref{fig:mw_observed} and \ref{fig:mw_unabs}, respectively.
Thanks to adding the UVOIR data, the 
parameters for the BB$_{\rm cold}$ component are much better constrained than for the X-ray fits alone. Contrary to the UVOIR-only fit, which only provides a brightness temperature $\propto (d/R)^2$, the best-fit temperatures in the joint fits are derived from both R-J and Wien parts of the BB spectra. Therefore, we can measure the $R/d$ ratio and the temperatures separately, despite a correlation between these parameters. 

The parameters of the BB$_{\rm hot}$ component and the absorption features in the joint X-ray + UVOIR fits are close to those obtained from the X-ray fits with the 1BBnG models: $kT_{\rm hot}\approx 105$--106 eV, $R_{\rm hot}/d_{260}\approx 1.5$--1.6 km, $E_{c1}\approx 0.73$--0.75 keV, $E_{c2}\approx 0.39$--0.41 keV, etc.
The ``cold radius'' values, $R_{\rm cold}/d_{260}\approx 5.5$--7 km,
are much lower while the ``cold temperature'' values, $kT_{\rm cold}\approx 30$--50 eV, values are substantially higher than for the 2BBnG models.
We see from Figures \ref{fig:mw_observed} and \ref{fig:mw_unabs} that the FUV emission is dominated by the BB$_{\rm cold}$ component, but both the BB$_{\rm hot}$ and the PL components also contribute, at $\sim 10\%$ level each.
The slope  
$-0.9\lesssim \alpha\lesssim -0.6$ of the UVOIR PL component, which dominates in the IR and optical ranges, is very close to that obtained from the UVOIR-only fit with the PL+BB model. 
Thus, we have shown that the phenomenological 2BBnG+PL models for the UVOIR + X-ray spectrum provide  
satisfactory fits with reasonable parameter values, better constrained than it is possible for separate UVOIR and X-ray data sets. The 2BB2G+PL model fits somewhat better than the 2BB1G+PL model, which can be considered as a justification for the additional spectral feature at $\approx 0.4$ keV. 

\subsection{Proper motion of J2143}
\label{sec:propmot}
Comparison of our and \citetalias{Kaplan2011}'s ACS/WFC F475W images, carried out in the same observational setup on MJD 60448.9 and MJD 55335.9, allows one to measure the proper motion of J2143  
with minimal distortion effects and good astrometric boresight correction possibility.
For the latter, we use the \emph{Gaia} data release 3, DR3 \citep{gaia2023,gaia2016} to identify 18 
suitable reference sources in the field of view. We applied the catalog proper motions to calculate the positions of the selected \emph{Gaia} DR3 stars at the two epochs. Using the Graphical Astronomy and Image Analysis Tool \citep{Currie2014}, we obtain a fit of the image astrometry with the proper-motion corrected \emph{Gaia} DR3 catalog positions.

The root mean square of the respective fit residuals is used as the systematic astrometric error (with respect to the common \emph{Gaia} DR3 reference frame). 
Centroid errors for the pulsar positions are obtained using FWHM values from 
2D Gaussian fits, and the peak signal-to-noise ratio of the detections according to 
Equation (1) in \citet{Reid1988}. 

The two HST positions in Table~\ref{tab:positions}
 are separated by $80.8 \pm 7.7$\,mas.
The proper motion components along R.A.\ 
and decl.\ 
are\footnote{Note that some authors use notations $\mu_{\alpha} \cos\delta$ and $\mu_l\cos b$ instead of $\mu_{\alpha}$ and $\mu_l$.}
\begin{equation}
\mu_\alpha = 4.3 \pm 0.5\,{\rm \,mas/yr },\quad \mu_\delta = -3.8 \pm 0.5\,{\rm mas/yr} . 
\end{equation}
This is equivalent to 
$\mu=5.8 \pm 0.5$\,mas\,yr$^{-1}$ 
with a position angle ${\rm P.A} = 131.3^\circ \pm  5.5 ^\circ$ (East of North; see Figure \ref{fig:HST_images}).

In Galactic coordinates, the proper motion vector is
$\mu_l = -0.2 \pm 0.5 $ mas yr$^{-1}$, $\mu_b= -5.8 \pm 0.5$ mas yr$^{-1}$,
i.e., J2143 (Galactic latitude of $b= -33.1^\circ$) is moving away from the Galactic plane.
At a fiducial distance of 260 pc, this proper motion corresponds to a transverse velocity of $7.1 \pm 0.7$\,km\,s$^{-1}$.

\begin{deluxetable*}{lcccllcc}
\label{tab:positions}
\tablecaption{Pulsar astrometry for the ACS/WFC F475W data} 
\tablehead{
 \colhead{epoch} & \colhead{pixel scale} &  \colhead{$\sigma^{\rm radial}_{\rm bsc}$} & \colhead{$\sigma^{\rm radial}_{\rm cent}$} &   
 \colhead{$\alpha-\alpha_0$} & \colhead{$\delta -\delta_0$} & \colhead{$\sigma^{\rm radial}$}& \colhead{$\sigma^{\rm 1D}$}\\  
 \colhead{} & \colhead{mas/pix} &  \colhead{mas} & \colhead{mas} & 
 \colhead{ } & \colhead{ } &\colhead{mas}  &\colhead{mas}   
}
\startdata
2010.4 &  50 &  6.9 & 0.4 & $3\fs39104$ & $17\farcs4268$ & 6.9 & 4.5\\
2024.4 &  50 &  9.4 & 0.9 & $3\fs39512$ & $17\farcs3735$ & 9.4 & 6.2
\enddata
\tablecomments{
The fit to the \emph{Gaia} DR3 counterparts of 18 field stars resulted in pulsar's R.A. $\alpha$ and decl.\ $\delta$, counted from $\alpha_0=21^{\rm h}\, 43^{\rm m}\,00^{\rm s}$ and $\delta_0=6^\circ\, 54\arcmin\, 00\arcsec$. 
The systematic radial uncertainty $\sigma^{\rm radial}_{\rm bsc}$ is the uncertainty of the boresight correction with the 18 stars, 
 $\sigma^{\rm radial}_{\rm cent}$ indicates the centroiding radial uncertainty, $\sigma^{\rm radial}$ is the total radial uncertainty, $\sigma^{\rm 1D}$ is the uncertainty in one coordinate component. {All the uncertainties are at the 68\% confidence level.}
}
\end{deluxetable*}

\section{Discussion} \label{sec:discussion}
Our HST observations have shown that the observed broadband UVOIR spectrum of J2143 
can be described by a sum of a thermal component,
dominating at UV wavelengths,
and a PL component, 
dominating in the NIR-optical (see Figure~\ref{fig:uvoir_spectrum}). Such a two-component UVOIR spectrum strongly resembles those of middle-aged RPPs (see 
\citealt{Abramkin2025}, and references therein), but it has not been clearly seen in XTINSs so far.
Below we discuss the thermal and PL components, taking the contemporaneous X-ray data into account.

\subsection{Nonthermal UVOIR emission}
The HST observations of J2143 in 5 spectral bands allowed us to discover a 
component 
with a PL-like spectrum, $f_\nu = f_0 (\nu/\nu_0)^\alpha$,  $f_0=40$--50 nJy 
at $\nu_0=10^{15}$ Hz, $-\alpha=0.6$--1.0. 
It resembles the nonthermal components
detected in UVOIR emission of several middle-aged RPPs (see Table 4 and Figure 8 in \citealt{Abramkin2025}),
with spectral slopes 
$-0.5\lesssim\alpha\lesssim -0.1$ and
typical luminosities $L^{\rm nonth}_{\rm 1-10\,eV} \equiv 4\pi d^2 F^{\rm nonth}_{\rm 1-10\,eV} \sim (2$--$7)\times 10^{28}$ erg s$^{-1}$, where $F_{\rm 1-10\,eV}^{\rm nonth}$ is the unabsorbed nonthermal energy flux in the 1--10 eV band.
Those components are usually interpreted as emission from relativistic particles 
in pulsar magnetospheres, powered by the pulsar's rotational energy loss rate $\dot{E}$.
The fraction of $\dot{E}$ converted into the UVOIR emission is characterized by
the efficiency $\eta_{\rm 1-10\,eV}\equiv L^{\rm nonth}_{\rm 1-10\,eV}/\Edot \sim (0.5$--$5)\times 10^{-6}$.

In the case of J2143, the 
UVOIR PL flux, 
luminosity and efficiency are 
\begin{eqnarray}
    F^{\rm PL}_{\rm 1-10\,eV} &=& (8-11)\times 10^{-16}\,\,{\rm erg}\,\,{\rm cm}^{-2}\,\,{\rm s}^{-1}, \\
    L^{\rm PL}_{\rm 1-10\,eV} & = & (6-9)\times 10^{27} d_{260}^2\,\, {\rm erg}\,\, {\rm s}^{-1},\\
    \eta^{\rm PL}_{\rm 1-10\,eV} & = & (3-5)\times 10^{-3} d_{260}^2\,,
\end{eqnarray}
i.e., the luminosity of J2143 is 
lower but the efficiency much higher than 
those of middle-aged RPPs.

Unlike RPPs, J2143 does not exhibit nonthermal emission in X-rays and $\gamma$-rays. 
\citet{Dessert2020} derived a flux upper limit of $7.4 \times 10^{-16}\,\,{\rm erg}\,\,{\rm cm}^{-2}\,\,{\rm s}^{-1}$ in the energy range 2--4\,keV,
which is below the extrapolation of the UVOIR PL with the same slope.
It is possible that the nonthermal $f_\nu$ spectrum 
steepens with increasing energy, as we observe in middle-aged RPPs, 
but an X-ray detection 
would be difficult 
even for very deep observations.

The differences between the alleged magnetospheric emission of J2143 and middle-aged RPPs 
could be associated with the longer period and lower spin-down power of the former, which may
lead to a smaller number of particles accelerated to high energies and a larger number of particles (and emitted photons) of lower energies. 
However, the lower spin-down power alone cannot be the cause of J2143's missing non-thermal X-ray and radio emission. Radio efficiency actually increases with lower spin-down power \citep{Posselt2023a}, and very old pulsars with comparable low spin-down power exhibit detectable nonthermal X-ray emission components \citep{Posselt2012}. 
On the other hand, the effect of the long period (e.g., on the Goldreich-Julian charge density) 
could be counterbalanced by the stronger magnetic field which should facilitate particle acceleration.
Although the actual emission mechanism in the UVOIR range remains unknown,  
it is possible that J2143 is a high-B, low-power cousin of ordinary middle-age RPPs whose non-thermal X-ray emission is simply below our detection threshold. 
%
The lack of $\gamma$-ray emission from J2143 can 
be explained by the low spin-down power
while the apparent lack 
of observable radio emission \citep{Kondratiev2009} could be due to unfavorable orientations of its rotation and magnetic axes or due to extreme faintness of the radio emission.

We note that in some respects the J2143's UVOIR PL component resembles that detected with JWST 
from the magnetar 4U\,0142+61, with the slope $\alpha=-0.96\pm0.02$ \citep{Hare2024}. If that emission is powered by the  
NS rotation energy loss, then the IR efficiency is as high as $\eta_{1.4-11\,\mu{\rm m}} \approx 0.6 (d/3.6\,{\rm kpc})^2$, which corresponds to $\eta_{\rm 1-10\,eV}\approx 0.7 (d/3.6\,{\rm kpc})^2$ if 
the IR-optical spectrum
is extrapolated to the UV range.
Thus, in terms of efficiency, J2143 could be an intermediate object between usual RPPs and magnetars. We caution, however, that 
the magnetar's UVOIR emission can be powered by a mechanism different from the rotation energy loss, in which case the above-defined efficiency becomes irrelevant.

In principle, the 
PL UVOIR component in the J2143 emission might be due to an 
unusual 
PWN that is not seen  
outside of the UVOIR range. A support for this interpretation might be provided by a confirmation of the extended emission around the point source, possibly seen in the two NIR filters (see Section \ref{sec:extended}). Such an explanation could be verified observationally by a search for UVOIR pulsations\footnote{Note that the nonthermal UVOIR components of at least some RPPs are pulsed \citep{Kargaltsev2007}, which likely excludes the PWN interpretation of the nonthermal component for those pulsars.}, but such a 
search  
is hardly feasible with the currently available instruments 
because of the
faintness of J2143 in the UVOIR. 

Yet another possibility might be that the UVOIR PL component 
$f_\nu \propto \nu^\alpha$ is 
emitted from a fallback 
disk with the temperature decreasing outward as $T\propto r^{-\beta}$, where $\beta = 2/(3-\alpha)$ (e.g., $\beta = 0.526$ for $\alpha = -0.8$)
-- see, for instance, \citet{Hare2024, Ertan2017}. 
This hypothesis
can be tested with IR observations at longer wavelengths.

To establish the true nature of the discovered 
PL UVOIR emission, 
it would be, of course, very important to understand 
how common such emission is among XTINSs (and perhaps high-B RPPs with similar properties). Indirect evidence for such emission is provided by the results of \citetalias{Kaplan2011}, who found
spectral slopes different from the R-J slope $\alpha=2$ in HST observations of 5 XTINSs (including J2143) in 2 spectral bands. A 
hint of nonthermal emission was also mentioned by \citet{Kaplan2003} in connection with the optical-UV spectrum of the peculiar XTINS RX\,J0720.4--3125, but other interpretations were not ruled out because of lack of deep (N)IR observations.
Finally, the flux densities of 
J0806, measured by \citetalias{Kaplan2011} in the 
F140LP and F475W bands, and by \citet{Posselt2018} in the F160W band,
are consistent with the spectrum being the sum of thermal and PL components, but it is hard to separate contributions to the NIR flux of the point source and the detected small-scale extended emission.
Thus, 
the presence of the UVOIR PL component in other XTINSs 
is currently unclear.

Regarding the nonthermal X-ray component, the only XTINS from which it has been apparently detected is the brightest 
J1856\footnote{For a second source, RX J0420.0--5022, there seems to be some excess which could be explained either with a blackbody or a PL component \citep{DeGrandis2022, Yoneyama2019}.}, which has been extensively studied with several X-ray observatories.
\citet{DeGrandis2022} collected 1.43 Ms of clean XMM-Newton and NICER data, fit the X-ray spectrum
with a 2BB+PL model, and found the PL component with a photon index $\Gamma=1.4^{+0.5}_{-0.4}$ (i.e., $\alpha=-0.4^{+0.4}_{-0.6}$), dominating at $E\gtrsim 1.5$ keV. The 2--8 keV flux of this component corresponds to the efficiency $\eta_{\rm 2-8\,keV}\sim 10^{-3}$. If we assume a similar X-ray efficiency for J2143, then its nonthermal flux would be
$F_{\rm 2-8\,keV}\sim 2.5\times 10^{-16}\eta_{-3} d_{260}^{-2}$ erg cm$^{-2}$ s$^{-1}$. 
This is below the constraint for the 4--8\,keV flux obtained by \citet{Dessert2020}.
For $\alpha = -0.4$, it corresponds to 
$f_\nu \sim 3\times 10^{-11}\eta_{-3} d_{260}^{-2}$ Jy and $\nu f_\nu \sim 0.8\times 10^{-16}\eta_{-3} d_{260}^{-2}$ erg cm$^{-2}$ s$^{-1}$ at $E=1$ keV. At $\eta_{-3}=1$ and $d_{260}=1$, these values are below the frames of Figures \ref{fig:mw_observed} and \ref{fig:mw_unabs}, i.e., such a nonthermal component is hardly detectable not only from J2143 but also from other XTINSs. 
It means that 
J1856 remains the only XTINS so far for which connection of the UVOIR and X-ray nonthermal emission can be studied, but there have been no deep observations of this object at $\lambda \gtrsim 6000$\,\AA.

\subsection{Thermal emission}
Fitting the UVOIR spectrum of J2143 with the PL + BB model (see Equation (\ref{eq:PL+BB_model}) and Figure \ref{fig:uvoir_spectrum}), we estimated the brightness temperature in the UV range: $kT_{\rm uv} \sim (15$--$25)(d_{260}/R_{10})^2$ eV, where $R_{10}$
is an equivalent  sphere radius in the UV range, $R_{\rm uv}$, in units of 10 km. Note that even if the UV emission emerges from a uniformly heated NS surface,  $R_{\rm uv}$ can be different from the NS radius because thermal emission may differ from the pure BB emission: 
$B_\nu(T) \to \varepsilon_\nu B_\nu(T)$ (with $\varepsilon_\nu \lesssim 1$),
which corresponds to $R_{\rm uv} \sim \varepsilon_{\rm uv}^{1/2} R_{\rm NS}$, where $\varepsilon_{\rm uv}$ is a characteristic normalized emissivity in the UV range. Because 
the actual values of $\varepsilon_{\rm uv}R_{\rm NS}/d$ are unknown, we cannot reliably estimate the NS surface temperature from UVOIR observations alone.

The temperature can be better constrained if we include X-ray data in the analysis,
which allows one to measure the (color) 
temperature and the $R/d$ ratio separately. 
Fitting the X-ray spectrum with a BB model with one or two absorption features (1BB1G and 1BB2G models in Table \ref{tab:xmm_spectral_vert}), we find the color temperature $kT_{\rm hot}\approx 106$ eV and the equivalent sphere radius $R_{\rm hot}\approx 1.5 d_{260}$ km. Extrapolation of this model to the UV range gives a flux that is a factor of $\sim 10$ lower than observed, i.e., the multiwavelength thermal spectrum cannot be fitted with a single-temperature BB model, similar to other XTINSs.
This apparent discrepancy can be caused by two factors -- nonuniformity of NS surface temperature and the difference of the real thermal spectrum from the BB spectrum.

The temperature nonuniformity can be due to
 the anisotropic heat transfer from the hot NS interiors in the strong magnetic field of XTINSs \citep{Greenstein1983}.
The temperature distribution depends on the unknown geometry of the NS magnetic field (e.g., \citealt{Pons2009}), but at least for some geometries the thermal emission spectrum can be approximated by a sum of two BB spectra \citep{Yakovlev2021}. Such models have been used for spectral fits of XTINSs (e.g., \citealt{Braje2002, Burwitz2003}) and RPPs (\citealt{Abramkin2025}, and references therein).

Applying this approach to the combined UVOIR and X-ray data, we fit the multiwavelength spectrum with the 2BB1G+PL and 2BB2G+PL models, assuming that the colder BB describes both the soft X-ray and UV parts of the spectrum. 
The better fit with the 2BB2G+PL model yields 
$kT_{\rm cold}\approx 40$--50 eV, 
$R_{\rm cold}\sim 6 d_{260}$ km,
and $L_{\rm cold} \sim 2\times 10^{31} d_{260}^2$ erg s$^{-1}$.
The estimated $kT_{\rm cold}$ 
is within the range of 40--70 eV 
found for these quantities for RPPs of similar ages (see Table 4 in \citealt{Abramkin2025}). 
It is slightly lower than $kT_{\rm uv}\approx 60$ eV, which we would obtain from the BB+PL fit of the UVOIR spectrum at $R_{\rm uv} = R_{\rm cold}$ (i.e., $R_{10} = 0.6 d_{260}$) 
and $A_V=0.12$ (see Figures \ref{fig:uvoir_spectrum} and \ref{fig:conf_contours}). The reason for this is the additional contribution from the
R-J tail of BB$_{\rm hot}$ in the F140LP band, $\sim 10\%$ of the total flux in that band, comparable to the contribution of the PL component (see Figures \ref{fig:mw_observed} and \ref{fig:mw_unabs}). 
The radius $R_{\rm cold}$ is smaller than the expected NS radius unless the target is 
unrealistically distant, $d\sim 600$--700 pc; this suggests a reduced  mean emissivity in the 
FUV through soft X-ray
range 
 and a surface emission spectrum different from the pure BB.
The bolometric luminosity of the BB$_{\rm cold}$ component is comparable to those of middle-aged RPPs that also have $R_{\rm cold}$ smaller than the expected NS radii.

Unlike middle-aged RPPs in which the BB$_{\rm hot}$ component usually dominates in the $\sim 0.5$--1.5 keV band, 
for J2143 it dominates in the entire soft X-ray range, 0.2--2 keV, where the target was detected with XMM-Newton (see Figure \ref{fig:mw_unabs}). 
The ratio $T_{\rm hot}/T_{\rm cold} \approx 2.2$
is typical for middle-aged RPPs,
but the ratio 
$R_{\rm hot}/R_{\rm cold} \sim 0.28$ is a factor of 3--7 larger than $R_{\rm hot}/R_{\rm cold} \sim 0.04$--0.08\footnote{Here we exclude Geminga from the set of middle-aged RPPs because the parameters of its faint hot component are very uncertain.} for middle-aged RPPs. 
Contrary to middle-aged RPPs, the luminosity $L_{\rm hot}\sim 
 4\times 10^{31} d_{260}^2$ erg s$^{-1}$ exceeds $L_{\rm cold}$.

Moreover, $L_{\rm hot}$ exceeds the rotation energy loss rate, $\dot{E}=2.0\times 10^{30}$ erg s$^{-1}$, at any reasonable distances. 
It 
means that the 
hot region of the J2143's surface is due to the anisotropic heat transfer rather than to the heating of the polar caps by relativistic particles precipitating from the magnetosphere. 

We should not forget that
the BB$_{\rm cold}$ + BB$_{\rm hot}$ model is 
a simplified description of thermal emission from an NS surface, whose true 
nature is not fully certain yet. For instance, some properties of the multiwavelength XTINS emission (such as the ``optical excess'' -- see Section \ref{sec:xtins}) could be explained by the presence of a gaseous
hydrogen or helium  atmosphere in the outer NS layers \citep{Pavlov1996}. However, the available models of 
at least 
optically thick 
light-element atmospheres 
cannot explain all the observed properties of XTINS and middle-aged RPPs. Even if a fit is statistically acceptable, it requires an unrealistically large $R/d$ ratio or/and a magnetic field inconsistent with other estimates, and it does not explain the observed X-ray absorption features (e.g., \citealt{Arumugasamy2018,Malacaria2019,Vahdat2024,Abramkin2025}). 
Models of optically thin hydrogen atmosphere above the solid NS surface 
can explain the multiwavelength spectra of some XTINS
\citep{Ho2007}, 
but they require a narrow range of atmosphere thickness, i.e., 
a fine-tuning of model parameters.

A more viable hypothesis is that the surfaces of NSs with high magnetic fields are 
in a condensed (solid or liquid) state at the relatively low temperatures of XTINSs and not-too-young RPPs (e.g, \citealt{Medin2007}), likely with a nonuniform surface temperature distribution. Thermal X-ray spectra, 
emitted by condensed NS surfaces, resemble BBs with a reduced local emissivity (e.g., \citealt{Perez-Azorin2005,Perez-Azorin2006}). 
In the X-ray range the frequency dependence of 
the local emissivity is rather complicated, with broad absorption features whose central frequencies and widths depend on direction and magnitude of the local magnetic field. These features are strongly smoothed by the the surface magnetic field nonuniformity  and NS rotation, 
so that the X-ray spectrum resembles a Planck spectrum with a reduced emissivity:
$f_\nu^{\rm unabs} \sim \pi (R_{\rm NS}/d)^2  \varepsilon_X B_\nu(T)$, 
i.e., 
the apparent radius is $\sim \varepsilon_X^{1/2} R_{\rm NS}$, $\varepsilon_X < 1$. 
Our assumption in the 2BBnG fits that the same radius $R_{\rm cold}$ is associated with both the X-ray and UV parts of the spectrum implies $\varepsilon_X \approx \varepsilon_{\rm uv}$. Such a relationship holds for the free-ions model of the condensed surface, while $\varepsilon_{\rm uv} \ll \varepsilon_X$ for the fixed-ions model \citep{vanAdelsberg2005}. 
The absorption spectral features at 0.74\,keV and (perhaps) 0.4\,keV 
could be related to the two depressions in the condensed surface spectrum, whose positions depend on the magnetic field as well as density and chemical composition of the condensed matter. 

 The above discussion is based on phase-integrated spectra of the thermal component. More information, particularly about the temperature distribution over the NS surface, can be obtained from the X-ray pulse profile and phase-resolved spectroscopy. We defer this analysis to future work, which should include the previous XMM-Newton, NICER, and eROSITA data sets and use realistic models for the energy-dependent angular distributions and direction-dependent spectra for the condensed surface emission. Here we only mention that the three-peak structure of the 0.15--2 keV pulse profile, shown in Figure~\ref{fig:pulse_profile}, is not consistent with 
models of internal surface heating in an NS with a centered dipole magnetic field (e.g., \citealt{Geppert2006}, and references therein).
The relatively low pulsed fraction of $\approx 4\%$ (which is, however, higher than $\approx 1.2\%$ for J1856) could be associated, for instance, with the J2143's spin axis being close to the line of sight, but the viewing angle can hardly be estimated without a detailed modeling.

\subsection{Extended emission}
The characteristic size of $\sim 2''$ of the putative extended emission from J2143, detected in the two NIR filters, 
corresponds to $\sim 0.8\times 10^{16} d_{260}$ cm.  
The flux densities $f_\nu^{\rm ext} = 1.69\pm0.25$ and $0.88\pm 0.17$ $\mu$Jy in the F160W and F110W filters, respectively, 
are a factor 
$\sim 10$ larger than those of the compact object.
They are
consistent with a PL spectrum 
with the slope $\alpha^{\rm ext} =-2.23\pm 0.83$, substantially steeper than the nonthermal NIR-optical emission of the pulsar. The flux in the 1--2 $\mu$m band can be estimated as $F_{1-2\,\mu{\rm m}}^{\rm ext}\approx 2.1\times 10^{-15}$ erg cm$^{-2}$ s$^{-1}$, which corresponds to the luminosity $L_{1-2\,\mu{\rm m}}^{\rm ext}\approx 1.7\times 10^{28} d_{260}^2$ erg s$^{-1}$. If this emission is powered by the rotation energy loss of the pulsar, such luminosity corresponds to a very high NIR efficiency, $\eta_{1-2\,\mu{\rm m}}^{\rm ext} \equiv L_{1-2\,\mu{\rm m}}^{\rm ext}/\dot{E} \approx 8.6\times 10^{-3} d_{260}^2$.

Putative extended NIR emission has been detected from only one other XTINS, RX\,J0806.4--4123 (J0806 hereafter), with a size of $\sim 0\farcs8$ \citep{Posselt2018}.
Its flux density in F160W, the only HST NIR band in which it has been observed, was estimated as
$f_\nu^{\rm ext} \sim 0.2$--0.4 $\mu$Jy. 
The F160W flux, $F_{\rm F160W}^{\rm ext}\sim (4$--$8)\times 10^{-16}$ erg cm$^{-2}$ s$^{-1}$, corresponds to the luminosity $L_{\rm F160W}^{\rm ext} \sim (3$--$6)\times 10^{27} d_{250}^2$ erg s$^{-1}$.
With $\dot{E}=2.6\times 10^{29}$ erg s$^{-1}$ \citep{Posselt2024}, the J0806 efficiency in this filter is $\eta_{\rm F160W}^{\rm ext} = (1.2$--$2.3)\times 10^{-2} d_{250}^2$. Thus, if J2143's extended emission is real, we see that the properties of the extended NIR emission are similar in these two XTINSs, which  
hints at
a similar origin.

Possible interpretations of such extended NIR emission -- a NIR-only 
PWN or a resolved disk around the NS -- are discussed in detail by \citet{Posselt2018}. Of these two interpretations, the former looks more plausible because of the large size of the extended structure. However, in the case of J2143, the putative extended emission is projected onto a crowded region, which strongly hampers the data analysis. Both J2143 and J0806 were also detected in the Far-Infrared at $160\mu$m with the Herschel Space telescope, but for J2143 the peak of the $160\mu$m is further away from the pulsar position and there are several other nearby FIR sources  \citep{Posselt2014}. It should also be noted that the extinction in the directions of both XTINSs shows a steep step-like increase. If the two XTINSs are both in a region of increased interstellar particle density, that could explain the non-detection of these PWNe outside the NIR range.


\subsection{Inferences from the measured proper motion}
\begin{figure*}[t]
\hspace{0.5cm}
\includegraphics[height=6cm]{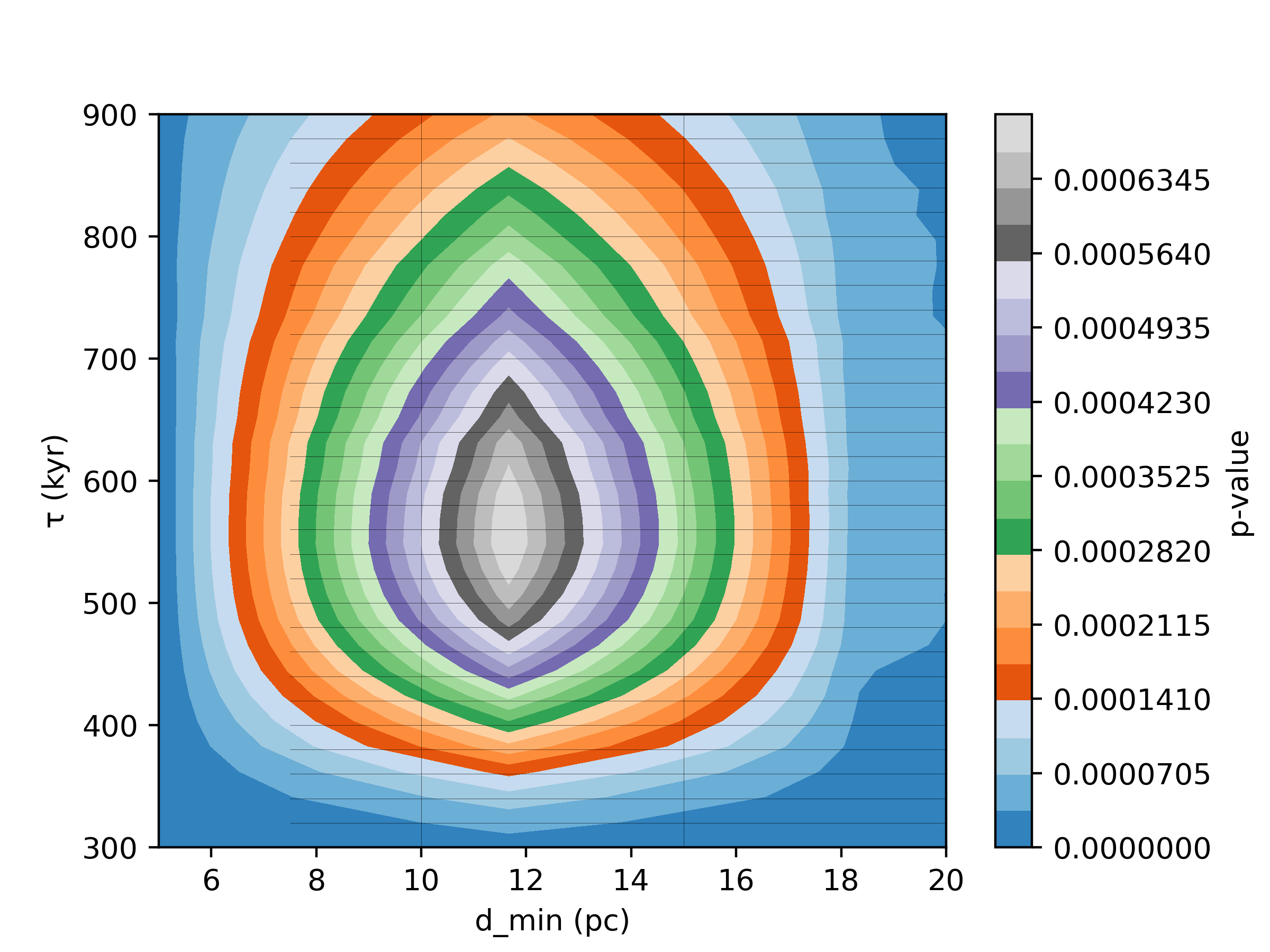}
\hspace{0.5cm}
\includegraphics[height=5.5cm]{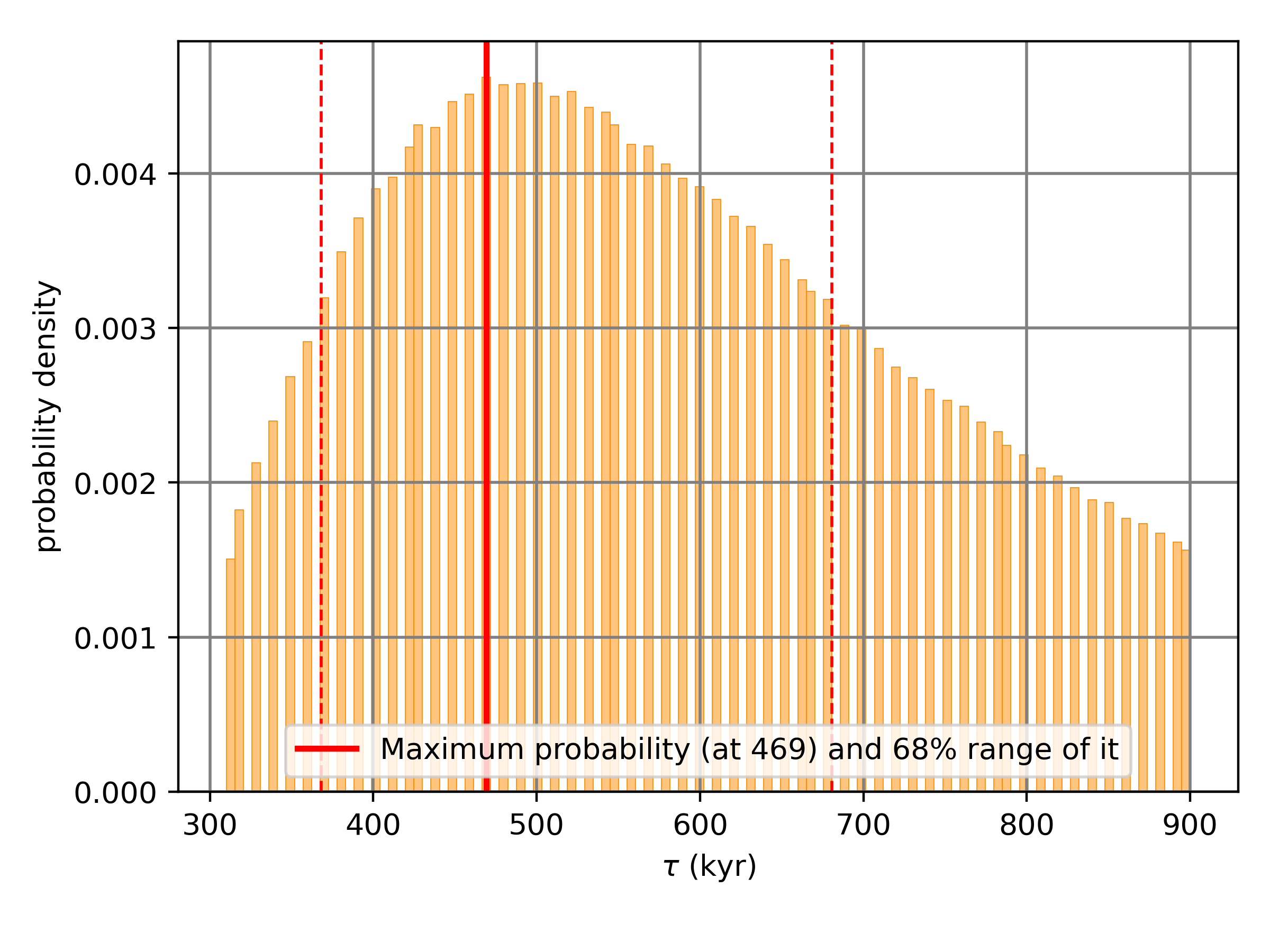}
\caption{$\beta$\,Pic-Cap stellar association as a possible birthplace of J2143. The left panel shows contours of the probability distribution for the encounter of J2143's trajectory with the stellar association; 
$d_{\rm min}$ is the smallest 
distance between the NS trajectory and the `center' of the association,
$\tau$ is the time interval between the closest encounter and the modern epoch.
The p-values at the color bar 
reflect the combined probability for the set of simulation input parameters (see \citet{Tetzlaff2010} for details). The image for the contour plot has bin sizes of 5\,pc in $d_{\rm min}$ and 20\,kyr in$\tau$. 
The right panel shows the age distribution of simulation sample from the left panel in a 1D histogram. The highest probability for an encounter between J2143 and $\beta$\,Pic-Cap is at $\sim 470$\,kyr.}
\label{fig:betapic}
\end{figure*}

The transverse velocity of J2143, $v_\perp=7.1\pm 0.7 d_{260}$ km\,s$^{-1} $, is one of the lowest measured for pulsars\footnote{Formally, it is the eighth lowest in the ATNF pulsar catalog, v2.6.5, 
but the transverse velocity values depend on the often poorly constrained distances.}.
Considering typical NS velocities \citep{Hobbs2005, Verbunt2017}, it is likely that the radial velocity is much higher. Following the same approach as \citet{Tetzlaff2010},
we simulate 2,000,000 different past trajectories of the pulsar in the Galactic potential to check for a possible origin of J2143 in a young stellar association\footnote{We used the list of young stellar association compiled by \citet{T2010Cat}}. For this, we assume a present distance of 260\,pc 
and use probability distributions for the proper motions, radial velocity, etc., similarly as described by \citet{Tetzlaff2010}.
At the characteristic age of the pulsar ($\sim 3$--4\,Myr), the probability of an overlap between the NS trajectory and any track of the 140 young star associations is very low.
Instead, the most probable and closest encounters between the NS and stellar associations 
took place less than 1\,Myr ago. This is in line with 
other XTINSs, whose kinetic ages are much smaller than characteristic ages. 
The $\beta$\,Pic-Cap stellar association (assumed size of 113\,pc and age of $8-34$\,Myr; \citealt{Tetzlaff2010}) 
has the highest probability of 
a possible trajectories crossing,
at a smallest separation of $\sim 14$\,pc from $\beta$\,Pic-Cap center $\sim 470$\,kyr ago (see Figure~\ref{fig:betapic}). 
Such a trajectory of J2143 would imply a today's radial velocity of 
360\,km\,s$^{-1}$,
with J2143 moving away from us. 
If the very close $\beta$\,Pic-Cap is indeed the birthplace of J2143, the XTINS's birth supernova happened relatively close ($\sim 70$\,pc) to the Sun.

\section{Conclusions}
\begin{itemize}
\item    The broadband UVOIR spectrum of J2143, obtained from our HST observations in 5 spectral bands, is dominated by a 
thermal (R-J) component in UV and a PL-like component 
at NIR-optical wavelengths, 
similar to the UVOIR spectra of middle-aged RPPs.
\item The flux density of the
NIR-optical component, corrected for the interstellar absorption with a plausible $V$-band extinction coefficient $A_V=0.12$, can be described by 
a  PL model 
$f_\nu^{\rm nonth}=f_0 (\nu/\nu_0)^\alpha$, with the slope $\alpha\sim -0.8$ 
and normalization $f_0 \sim 45$ nJy at $\nu_0=10^{15}$ Hz. 
The luminosity of this component in the UVOIR range is
$L^{\rm nonth}_{\rm 1-10\,eV} \sim  
7\times 10^{27}$
erg s$^{-1}$ at the likely distance $d=260$ pc, 
estimated from 3D extinction models and X-ray absorption.
If powered by the pulsar's rotation energy loss, this luminosity corresponds to a very high efficiency of converting the spin-down power to UVOIR radiation: $\eta_{\rm 1-10\,eV} \sim 4\times 10^{-3}$.
%
\item 
The thermal UV component, $f_\nu^{\rm therm}\propto \nu^2$, is a R-J tail of a Planck-like thermal spectrum. The UVOIR observations provide an estimate of brightness temperature, $kT_{\rm uv}\sim 20 d_{260}^2 (R_{\rm uv}/10\,{\rm km})^{-2}$ eV, rather uncertain because of the unknown effective 
size of the UV-emitting region.
The temperature is better constrained from the joint analysis of the UVOIR + X-ray spectrum, which can be fitted with the sum BB$_{\rm hot}$ + BB$_{\rm cold}$, with $kT_{\rm hot}\sim 106$ eV, $R_{\rm hot}\sim 1.5 d_{260}$ km, 
$L_{\rm hot}\sim 4\times 10^{31} d_{260}^2$, 
and $kT_{\rm cold} \sim 45$ eV, $R_{\rm cold}\sim 6 d_{260}$ km,
$L_{\rm cold}\sim 2\times 10^{31} d_{260}^2$ erg s$^{-1}$. 
Both the `hot' and `cold' bolometric luminosities exceed the spin-down power $\dot{E}=2.0\times 10^{30}$ erg s$^{-1}$, which 
supports the common assumption for XTINSs 
that the thermal emission is due to internal heating mechanisms rather than external heating by (magnetospheric) relativistic particles.
\item 
The images in the two NIR bands show hints of extended emission, with a $\sim 2''$ size and flux densities $f_\nu^{\rm ext} \sim 1.7$ $\mu$Jy and 0.9 $\mu$Jy at $\lambda= 1.54$ and 1.15 $\mu$m, respectively.
They correspond to the NIR luminosity $L_{1-2\,\mu{\rm m}}^{\rm ext} \sim 2\times 10^{28} d_{260}^2$ erg s$^{-1}$.
If confirmed,
the extended emission can be a peculiar infrared 
PWN with a very high efficiency, $\eta_{1-2\,\mu{\rm m}} 
\sim 
0.01 d_{260}^2$,
perhaps similar to the extended emission around 
J0806.
%
\item 
The measured proper motion of J2143, $\mu\sim 6$ mas yr$^{-1}$, corresponds to a low transverse velocity $v_\perp \sim 7 d_{260}$ km s$^{-1}$).
This XTINS could have been born $\sim 0.5$\,Myr ago very close to the Sun, in the stellar association $\beta$\,Pic-Cap.
Similar to the the other XTINSs for which a `kinematic age' has been estimated, J2143 seems to be substantially younger than its characteristic age of 3.6 Myr. 
\item 
Our spectral and timing analyses of the X-ray emission, obtained in a contemporaneous observation with the XMM-Newton Observatory, did not show significant differences 
with previous XMM-Newton and NICER observations of J2143.
\item 
We suggest that 
other XTINSs have similar UVOIR spectra, comprised of thermal and 
PL components, dominating in the UV and optical-IR ranges, respectively. It could be confirmed by more comprehensive UVOIR observations of these objects.
\item 
The UVOIR + X-ray 
spectrum of J2143 shows many properties similar to those of RPPs of comparable ages. The differences between these two types of objects 
could be 
due to higher magnetic fields of newborn XTINSs, which led to additional heating, faster slow-down at their young ages, longer periods, and 
a shift of the nonthermal spectrum to lower photon energies.
\end{itemize}

\begin{acknowledgements}
We appreciate the help by Oleg Kargaltsev in preparation of the observation proposal and the useful advice by Jay Anderson on the observational setup for the NIR observations. We thank Norbert Schartel for granting the TOO XMM-Newton observation contemporaneous with the HST observation.
We are indebted to Yura Shibanov for sharing the results of spectral analysis of previous X-ray observations of J2143. 
We especially thank Rosine Lallement and 
Gordian Edenhofer for helpful discussions and providing extinction maps in the direction toward J2143, which allowed us to 
obtain a reliable estimate for the distance to this object and the corresponding extinction.
We warmly thank Nina Tetzlaff, Kieran Moore, and Arnas Matulaitis for their help with the birthplace simulation code.

 Support for program \#17476 was provided by NASA through
a grant from the Space Telescope Science Institute, which is
operated by the Association of Universities for Research in
Astronomy, Inc., under NASA contract NAS 5-26555.
Some of the data presented in this paper were obtained from
the Mikulski Archive for Space Telescopes (MAST) at the
Space Telescope Science Institute. 

This research has used data, tools or materials developed as part of the EXPLORE project that has received funding from the European Union's Horizon 2020 research and innovation programme under grant agreement No 101004214.
This work has made use of data from the European Space Agency (ESA) mission
{\it Gaia} (\url{https://www.cosmos.esa.int/gaia}), processed by the {\it Gaia}
Data Processing and Analysis Consortium (DPAC,
\url{https://www.cosmos.esa.int/web/gaia/dpac/consortium}). Funding for the DPAC
has been provided by national institutions, in particular the institutions
participating in the {\it Gaia} Multilateral Agreement.

 The HST data presented in this article were obtained from the Mikulski Archive for Space Telescopes (MAST) at the Space Telescope Science Institute. The specific observations analyzed can be accessed via \dataset[doi: 10.17909/9jev-5m35]{https://doi.org/10.17909/9jev-5m35}.

\end{acknowledgements}

\facilities{HST(ACS/SBC, ACS/WFC, WFC3/UVIS, WFC3/IR) and XMM-Newton(EPIC).}
\software{XSPEC \citep{Arnaud1996}, XMM SAS \citep{Gabriel2004}, Graphical Astronomy and Image Analysis Tool \citep{Currie2014}, G-Tomo \citep{Lallement2022},
Astropy \citep{AstropyCollaboration2022}, Matplotlib \citep{Hunter2007}.}

\appendix
\section{The flux in the region around the neutron star}
\label{extmeas}
\begin{figure*}[ht]
\includegraphics[width=18cm]{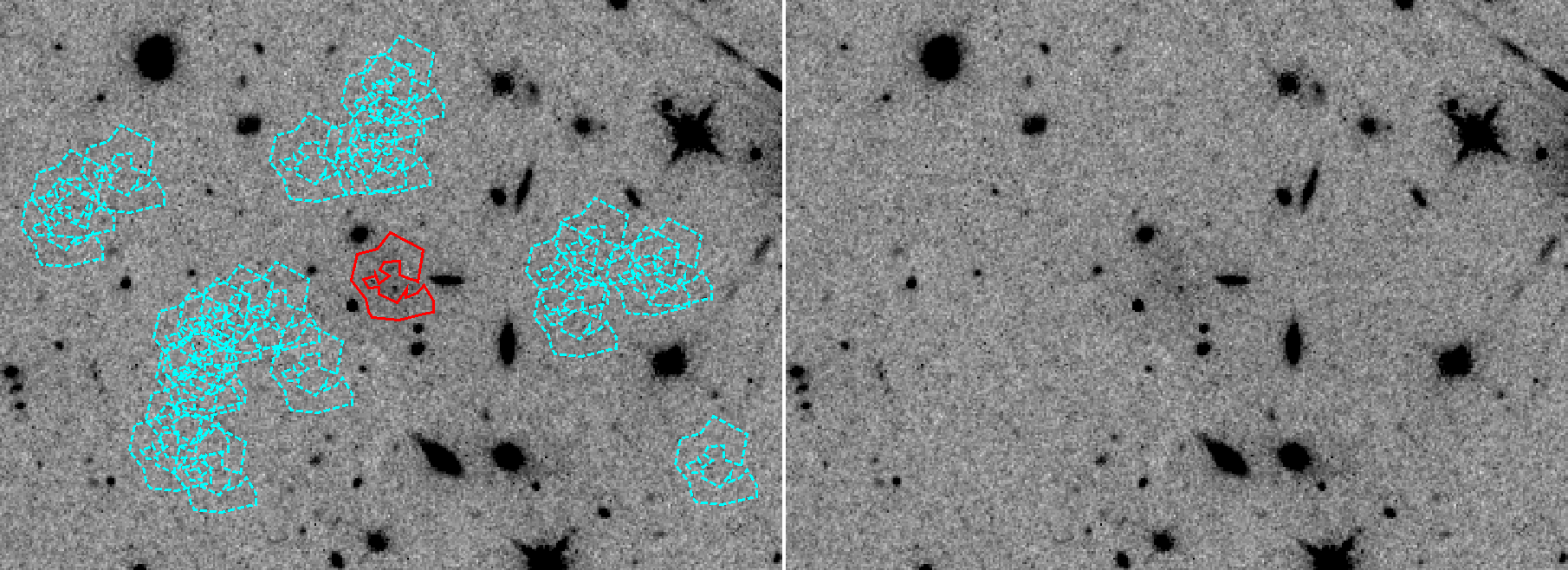}
\caption{The polygon apertures in the F160W image that were used to measure the apparent flux enhancement in the region around the NS (red). The used 26 background apertures with the same shape are marked with cyan dashed regions. The right panel shows the same image (about $46\arcsec \times 33\arcsec$, North up, East to the left) to allow an undisturbed view.
For a zoom-in of the source region, see Figure~\ref{fig:HST_images}.
\label{fig:exapertures}}
\end{figure*}

We noticed some possible flux enhancement around the target in the F160W image. It seems most prominent in an area with distance up to $2\arcsec$ from J2143. As the region is very crowded, there is the possibility of very faint background sources in that area. Aiming to avoid  such potential background sources, we define a polygon aperture (area of $\approx 13 \arcsec^2$) around the NS, see Figure~\ref{fig:exapertures}. We use the same polygon shape for 26 background apertures that allow us to measure the background flux and its error. We then use the same procedure as in Section~\ref{sec:photometry}, equation~(\ref{eq:netfluxerr}) to determine the net flux density in the polygon region, $f_{\rm F160W}^{\rm ext} = 1.69\pm 0.25$\,$\mu$Jy.
We use the same source aperture and procedure to measure the flux density in the F110W band, $f_{\rm F110W}^{\rm ext} = 0.88\pm 0.17$\,$\mu$Jy.

\bibliography{J2143}

@INPROCEEDINGS{Currie2014,
       author = {{Currie}, M.~J. and {Berry}, D.~S. and {Jenness}, T. and {Gibb}, A.~G. and
         {Bell}, G.~S. and {Draper}, P.~W.},
        title = "{Starlink Software in 2013}",
    booktitle = {Astronomical Data Analysis Software and Systems XXIII},
         year = 2014,
       editor = {{Manset}, N. and {Forshay}, P.},
       series = {Astronomical Society of the Pacific Conference Series},
       volume = {485},
        month = may,
        pages = {391},
       adsurl = {https://ui.adsabs.harvard.edu/abs/2014ASPC..485..391C}
}

@ARTICLE{Reid1988,
       author = {{Reid}, M.~J. and {Schneps}, M.~H. and {Moran}, J.~M. and
         {Gwinn}, C.~R. and {Genzel}, R. and {Downes}, D. and {Roennaeng}, B.},
        title = "{The Distance to the Center of the Galaxy: H 2O Maser Proper Motions in Sagittarius B2(N)}",
      journal = {\apj},
         year = 1988,
        month = jul,
       volume = {330},
        pages = {809},
          doi = {10.1086/166514},
       adsurl = {https://ui.adsabs.harvard.edu/abs/1988ApJ...330..809R}
}

@ARTICLE{gaia2023,
       author = {{Gaia Collaboration} and {Vallenari}, A. and {Brown}, A.~G.~A. and {Prusti}, T. and {de Bruijne}, J.~H.~J. and {Arenou}, F. and {Babusiaux}, C. and {Biermann}, M. and {Creevey}, O.~L. and {Ducourant}, C. and {Evans}, D.~W. and {Eyer}, L. and {Guerra}, R. and {Hutton}, A. and {Jordi}, C. and {Klioner}, S.~A. and {Lammers}, U.~L. and {Lindegren}, L. and {Luri}, X. and {Mignard}, F. and {Panem}, C. and {Pourbaix}, D. and {Randich}, S. and {Sartoretti}, P. and {Soubiran}, C. and {Tanga}, P. and {Walton}, N.~A. and {Bailer-Jones}, C.~A.~L. and {Bastian}, U. and {Drimmel}, R. and {Jansen}, F. and {Katz}, D. and {Lattanzi}, M.~G. and {van Leeuwen}, F. and {Bakker}, J. and {Cacciari}, C. and {Casta{\~n}eda}, J. and {De Angeli}, F. and {Fabricius}, C. and {Fouesneau}, M. and {Fr{\'e}mat}, Y. and {Galluccio}, L. and {Guerrier}, A. and {Heiter}, U. and {Masana}, E. and {Messineo}, R. and {Mowlavi}, N. and {Nicolas}, C. and {Nienartowicz}, K. and {Pailler}, F. and {Panuzzo}, P. and {Riclet}, F. and {Roux}, W. and {Seabroke}, G.~M. and {Sordo}, R. and {Th{\'e}venin}, F. and {Gracia-Abril}, G. and {Portell}, J. and {Teyssier}, D. and {Altmann}, M. and {Andrae}, R. and {Audard}, M. and {Bellas-Velidis}, I. and {Benson}, K. and {Berthier}, J. and {Blomme}, R. and {Burgess}, P.~W. and {Busonero}, D. and {Busso}, G. and {C{\'a}novas}, H. and {Carry}, B. and {Cellino}, A. and {Cheek}, N. and {Clementini}, G. and {Damerdji}, Y. and {Davidson}, M. and {de Teodoro}, P. and {Nu{\~n}ez Campos}, M. and {Delchambre}, L. and {Dell'Oro}, A. and {Esquej}, P. and {Fern{\'a}ndez-Hern{\'a}ndez}, J. and {Fraile}, E. and {Garabato}, D. and {Garc{\'\i}a-Lario}, P. and {Gosset}, E. and {Haigron}, R. and {Halbwachs}, J. -L. and {Hambly}, N.~C. and {Harrison}, D.~L. and {Hern{\'a}ndez}, J. and {Hestroffer}, D. and {Hodgkin}, S.~T. and {Holl}, B. and {Jan{\ss}en}, K. and {Jevardat de Fombelle}, G. and {Jordan}, S. and {Krone-Martins}, A. and {Lanzafame}, A.~C. and {L{\"o}ffler}, W. and {Marchal}, O. and {Marrese}, P.~M. and {Moitinho}, A. and {Muinonen}, K. and {Osborne}, P. and {Pancino}, E. and {Pauwels}, T. and {Recio-Blanco}, A. and {Reyl{\'e}}, C. and {Riello}, M. and {Rimoldini}, L. and {Roegiers}, T. and {Rybizki}, J. and {Sarro}, L.~M. and {Siopis}, C. and {Smith}, M. and {Sozzetti}, A. and {Utrilla}, E. and {van Leeuwen}, M. and {Abbas}, U. and {{\'A}brah{\'a}m}, P. and {Abreu Aramburu}, A. and {Aerts}, C. and {Aguado}, J.~J. and {Ajaj}, M. and {Aldea-Montero}, F. and {Altavilla}, G. and {{\'A}lvarez}, M.~A. and {Alves}, J. and {Anders}, F. and {Anderson}, R.~I. and {Anglada Varela}, E. and {Antoja}, T. and {Baines}, D. and {Baker}, S.~G. and {Balaguer-N{\'u}{\~n}ez}, L. and {Balbinot}, E. and {Balog}, Z. and {Barache}, C. and {Barbato}, D. and {Barros}, M. and {Barstow}, M.~A. and {Bartolom{\'e}}, S. and {Bassilana}, J. -L. and {Bauchet}, N. and {Becciani}, U. and {Bellazzini}, M. and {Berihuete}, A. and {Bernet}, M. and {Bertone}, S. and {Bianchi}, L. and {Binnenfeld}, A. and {Blanco-Cuaresma}, S. and {Blazere}, A. and {Boch}, T. and {Bombrun}, A. and {Bossini}, D. and {Bouquillon}, S. and {Bragaglia}, A. and {Bramante}, L. and {Breedt}, E. and {Bressan}, A. and {Brouillet}, N. and {Brugaletta}, E. and {Bucciarelli}, B. and {Burlacu}, A. and {Butkevich}, A.~G. and {Buzzi}, R. and {Caffau}, E. and {Cancelliere}, R. and {Cantat-Gaudin}, T. and {Carballo}, R. and {Carlucci}, T. and {Carnerero}, M.~I. and {Carrasco}, J.~M. and {Casamiquela}, L. and {Castellani}, M. and {Castro-Ginard}, A. and {Chaoul}, L. and {Charlot}, P. and {Chemin}, L. and {Chiaramida}, V. and {Chiavassa}, A. and {Chornay}, N. and {Comoretto}, G. and {Contursi}, G. and {Cooper}, W.~J. and {Cornez}, T. and {Cowell}, S. and {Crifo}, F. and {Cropper}, M. and {Crosta}, M. and {Crowley}, C. and {Dafonte}, C. and {Dapergolas}, A. and {David}, M. and {David}, P. and {de Laverny}, P. and {De Luise}, F. and {De March}, R.},
        title = "{Gaia Data Release 3. Summary of the content and survey properties}",
      journal = {\aap},
         year = 2023,
        month = jun,
       volume = {674},
          eid = {A1},
        pages = {A1},
          doi = {10.1051/0004-6361/202243940},
archivePrefix = {arXiv},
       eprint = {2208.00211},
 primaryClass = {astro-ph.GA},
       adsurl = {https://ui.adsabs.harvard.edu/abs/2023A&A...674A...1G}
}

@ARTICLE{gaia2016,
       author = {{Gaia Collaboration} and {Prusti}, T. and {de Bruijne}, J.~H.~J. and {Brown}, A.~G.~A. and {Vallenari}, A. and {Babusiaux}, C. and {Bailer-Jones}, C.~A.~L. and {Bastian}, U. and {Biermann}, M. and {Evans}, D.~W. and {Eyer}, L. and {Jansen}, F. and {Jordi}, C. and {Klioner}, S.~A. and {Lammers}, U. and {Lindegren}, L. and {Luri}, X. and {Mignard}, F. and {Milligan}, D.~J. and {Panem}, C. and {Poinsignon}, V. and {Pourbaix}, D. and {Randich}, S. and {Sarri}, G. and {Sartoretti}, P. and {Siddiqui}, H.~I. and {Soubiran}, C. and {Valette}, V. and {van Leeuwen}, F. and {Walton}, N.~A. and {Aerts}, C. and {Arenou}, F. and {Cropper}, M. and {Drimmel}, R. and {H{\o}g}, E. and {Katz}, D. and {Lattanzi}, M.~G. and {O'Mullane}, W. and {Grebel}, E.~K. and {Holland}, A.~D. and {Huc}, C. and {Passot}, X. and {Bramante}, L. and {Cacciari}, C. and {Casta{\~n}eda}, J. and {Chaoul}, L. and {Cheek}, N. and {De Angeli}, F. and {Fabricius}, C. and {Guerra}, R. and {Hern{\'a}ndez}, J. and {Jean-Antoine-Piccolo}, A. and {Masana}, E. and {Messineo}, R. and {Mowlavi}, N. and {Nienartowicz}, K. and {Ord{\'o}{\~n}ez-Blanco}, D. and {Panuzzo}, P. and {Portell}, J. and {Richards}, P.~J. and {Riello}, M. and {Seabroke}, G.~M. and {Tanga}, P. and {Th{\'e}venin}, F. and {Torra}, J. and {Els}, S.~G. and {Gracia-Abril}, G. and {Comoretto}, G. and {Garcia-Reinaldos}, M. and {Lock}, T. and {Mercier}, E. and {Altmann}, M. and {Andrae}, R. and {Astraatmadja}, T.~L. and {Bellas-Velidis}, I. and {Benson}, K. and {Berthier}, J. and {Blomme}, R. and {Busso}, G. and {Carry}, B. and {Cellino}, A. and {Clementini}, G. and {Cowell}, S. and {Creevey}, O. and {Cuypers}, J. and {Davidson}, M. and {De Ridder}, J. and {de Torres}, A. and {Delchambre}, L. and {Dell'Oro}, A. and {Ducourant}, C. and {Fr{\'e}mat}, Y. and {Garc{\'\i}a-Torres}, M. and {Gosset}, E. and {Halbwachs}, J. -L. and {Hambly}, N.~C. and {Harrison}, D.~L. and {Hauser}, M. and {Hestroffer}, D. and {Hodgkin}, S.~T. and {Huckle}, H.~E. and {Hutton}, A. and {Jasniewicz}, G. and {Jordan}, S. and {Kontizas}, M. and {Korn}, A.~J. and {Lanzafame}, A.~C. and {Manteiga}, M. and {Moitinho}, A. and {Muinonen}, K. and {Osinde}, J. and {Pancino}, E. and {Pauwels}, T. and {Petit}, J. -M. and {Recio-Blanco}, A. and {Robin}, A.~C. and {Sarro}, L.~M. and {Siopis}, C. and {Smith}, M. and {Smith}, K.~W. and {Sozzetti}, A. and {Thuillot}, W. and {van Reeven}, W. and {Viala}, Y. and {Abbas}, U. and {Abreu Aramburu}, A. and {Accart}, S. and {Aguado}, J.~J. and {Allan}, P.~M. and {Allasia}, W. and {Altavilla}, G. and {{\'A}lvarez}, M.~A. and {Alves}, J. and {Anderson}, R.~I. and {Andrei}, A.~H. and {Anglada Varela}, E. and {Antiche}, E. and {Antoja}, T. and {Ant{\'o}n}, S. and {Arcay}, B. and {Atzei}, A. and {Ayache}, L. and {Bach}, N. and {Baker}, S.~G. and {Balaguer-N{\'u}{\~n}ez}, L. and {Barache}, C. and {Barata}, C. and {Barbier}, A. and {Barblan}, F. and {Baroni}, M. and {Barrado y Navascu{\'e}s}, D. and {Barros}, M. and {Barstow}, M.~A. and {Becciani}, U. and {Bellazzini}, M. and {Bellei}, G. and {Bello Garc{\'\i}a}, A. and {Belokurov}, V. and {Bendjoya}, P. and {Berihuete}, A. and {Bianchi}, L. and {Bienaym{\'e}}, O. and {Billebaud}, F. and {Blagorodnova}, N. and {Blanco-Cuaresma}, S. and {Boch}, T. and {Bombrun}, A. and {Borrachero}, R. and {Bouquillon}, S. and {Bourda}, G. and {Bouy}, H. and {Bragaglia}, A. and {Breddels}, M.~A. and {Brouillet}, N. and {Br{\"u}semeister}, T. and {Bucciarelli}, B. and {Budnik}, F. and {Burgess}, P. and {Burgon}, R. and {Burlacu}, A. and {Busonero}, D. and {Buzzi}, R. and {Caffau}, E. and {Cambras}, J. and {Campbell}, H. and {Cancelliere}, R. and {Cantat-Gaudin}, T. and {Carlucci}, T. and {Carrasco}, J.~M. and {Castellani}, M. and {Charlot}, P. and {Charnas}, J. and {Charvet}, P. and {Chassat}, F. and {Chiavassa}, A. and {Clotet}, M. and {Cocozza}, G. and {Collins}, R.~S. and {Collins}, P. and {Costigan}, G.},
        title = "{The Gaia mission}",
      journal = {\aap},
         year = 2016,
        month = nov,
       volume = {595},
          eid = {A1},
        pages = {A1},
          doi = {10.1051/0004-6361/201629272},
archivePrefix = {arXiv},
       eprint = {1609.04153},
 primaryClass = {astro-ph.IM},
       adsurl = {https://ui.adsabs.harvard.edu/abs/2016A&A...595A...1G}
}

@MISC{Rodrigo2012,
       author = {{Rodrigo}, Carlos and {Solano}, Enrique and {Bayo}, Amelia},
        title = "{SVO Filter Profile Service Version 1.0}",
 howpublished = {IVOA Working Draft 15 October 2012},
         year = 2012,
        month = oct,
        pages = {1015},
          doi = {10.5479/ADS/bib/2012ivoa.rept.1015R},
       adsurl = {https://ui.adsabs.harvard.edu/abs/2012ivoa.rept.1015R}
}

@Article{Hunter2007,
  Author    = {Hunter, J. D.},
  Title     = {Matplotlib: A 2D graphics environment},
  Journal   = {Computing in Science \& Engineering},
  Volume    = {9},
  Number    = {3},
  Pages     = {90--95},
  publisher = {IEEE COMPUTER SOC},
  doi       = {10.1109/MCSE.2007.55},
  year      = 2007
}

@INPROCEEDINGS{Arnaud1996,
       author = {{Arnaud}, K.~A.},
        title = "{XSPEC: The First Ten Years}",
    booktitle = {Astronomical Data Analysis Software and Systems V},
         year = 1996,
       editor = {{Jacoby}, George H. and {Barnes}, Jeannette},
       series = {Astronomical Society of the Pacific Conference Series},
       volume = {101},
        month = jan,
        pages = {17},
       adsurl = {https://ui.adsabs.harvard.edu/abs/1996ASPC..101...17A}
}

@ARTICLE{Verbunt2017,
       author = {{Verbunt}, Frank and {Igoshev}, Andrei and {Cator}, Eric},
        title = "{The observed velocity distribution of young pulsars}",
      journal = {\aap},
         year = 2017,
        month = dec,
       volume = {608},
          eid = {A57},
        pages = {A57},
          doi = {10.1051/0004-6361/201731518},
archivePrefix = {arXiv},
       eprint = {1708.08281},
 primaryClass = {astro-ph.HE},
       adsurl = {https://ui.adsabs.harvard.edu/abs/2017A&A...608A..57V}
}

@ARTICLE{Hobbs2005,
       author = {{Hobbs}, G. and {Lorimer}, D.~R. and {Lyne}, A.~G. and {Kramer}, M.},
        title = "{A statistical study of 233 pulsar proper motions}",
      journal = {\mnras},
         year = 2005,
        month = jul,
       volume = {360},
       number = {3},
        pages = {974-992},
          doi = {10.1111/j.1365-2966.2005.09087.x},
archivePrefix = {arXiv},
       eprint = {astro-ph/0504584},
 primaryClass = {astro-ph},
       adsurl = {https://ui.adsabs.harvard.edu/abs/2005MNRAS.360..974H}
}

@dataset{T2010Cat,
       author = {{Tetzlaff}, N. and {Neuhaeuser}, R. and {Hohle}, M.~M. and {Maciejewski}, G.},
        title = "{VizieR Online Data Catalog: Kinematics of young associations/clusters (Tetzlaff+, 2010)}",
 howpublished = {VizieR On-line Data Catalog: J/MNRAS/402/2369. Originally published in: 2010MNRAS.402.2369T},
         year = 2010,
        month = mar,
          eid = {J/MNRAS/402/2369},
       adsurl = {https://ui.adsabs.harvard.edu/abs/2010yCat..74022369T}
}

@ARTICLE{Tetzlaff2010,
       author = {{Tetzlaff}, N. and {Neuh{\"a}user}, R. and {Hohle}, M.~M. and {Maciejewski}, G.},
        title = "{Identifying birth places of young isolated neutron stars}",
      journal = {\mnras},
         year = 2010,
        month = mar,
       volume = {402},
       number = {4},
        pages = {2369-2387},
          doi = {10.1111/j.1365-2966.2009.16093.x},
archivePrefix = {arXiv},
       eprint = {0911.4441},
 primaryClass = {astro-ph.GA},
       adsurl = {https://ui.adsabs.harvard.edu/abs/2010MNRAS.402.2369T}
}

@ARTICLE{Yoneyama2019,
       author = {{Yoneyama}, Tomokage and {Hayashida}, Kiyoshi and {Nakajima}, Hiroshi and {Matsumoto}, Hironori},
        title = "{Universal detection of high-temperature emission in X-ray isolated neutron stars}",
      journal = {\pasj},
         year = 2019,
        month = jan,
       volume = {71},
       number = {1},
          eid = {17},
        pages = {17},
          doi = {10.1093/pasj/psy135},
archivePrefix = {arXiv},
       eprint = {1811.11982},
 primaryClass = {astro-ph.HE},
       adsurl = {https://ui.adsabs.harvard.edu/abs/2019PASJ...71...17Y}
}

@ARTICLE{DeGrandis2022,
       author = {{De Grandis}, Davide and {Rigoselli}, Michela and {Mereghetti}, Sandro and {Younes}, George and {Pizzochero}, Pierre and {Taverna}, Roberto and {Tiengo}, Andrea and {Turolla}, Roberto and {Zane}, Silvia},
        title = "{Two decades of X-ray observations of the isolated neutron star RX J1856.5 - 3754: detection of thermal and non-thermal hard X-rays and refined spin-down measurement}",
      journal = {\mnras},
         year = 2022,
        month = nov,
       volume = {516},
       number = {4},
        pages = {4932-4941},
          doi = {10.1093/mnras/stac2587},
archivePrefix = {arXiv},
       eprint = {2209.03874},
 primaryClass = {astro-ph.HE},
       adsurl = {https://ui.adsabs.harvard.edu/abs/2022MNRAS.516.4932D}
}

@ARTICLE{Malacaria2019,
       author = {{Malacaria}, Christian and {Bogdanov}, Slavko and {Ho}, Wynn C.~G. and {Enoto}, Teruaki and {Ray}, Paul S. and {Arzoumanian}, Zaven and {Cazeau}, Thoniel and {Gendreau}, Keith C. and {Guillot}, Sebastien and {G{\"u}ver}, Tolga and {Jaisawal}, Gaurava K. and {Wolff}, Michael T. and {NICER Magnetar} and {Magnetospheres Team}},
        title = "{A Joint NICER and XMM-Newton View of the {\textquotedblleft}Magnificent{\textquotedblright} Thermally Emitting X-Ray Isolated Neutron Star RX J1605.3+3249}",
      journal = {\apj},
         year = 2019,
        month = aug,
       volume = {880},
       number = {2},
          eid = {74},
        pages = {74},
          doi = {10.3847/1538-4357/ab2875},
archivePrefix = {arXiv},
       eprint = {1906.02806},
 primaryClass = {astro-ph.HE},
       adsurl = {https://ui.adsabs.harvard.edu/abs/2019ApJ...880...74M}
}

@ARTICLE{Haberl2007,
       author = {{Haberl}, Frank},
        title = "{The magnificent seven: magnetic fields and surface temperature distributions}",
      journal = {\apss},
         year = 2007,
        month = apr,
       volume = {308},
       number = {1-4},
        pages = {181-190},
          doi = {10.1007/s10509-007-9342-x},
archivePrefix = {arXiv},
       eprint = {astro-ph/0609066},
 primaryClass = {astro-ph},
       adsurl = {https://ui.adsabs.harvard.edu/abs/2007Ap&SS.308..181H}
}

@ARTICLE{Hohle2012,
       author = {{Hohle}, M.~M. and {Haberl}, F. and {Vink}, J. and {de Vries}, C.~P. and {Turolla}, R. and {Zane}, S. and {M{\'e}ndez}, M.},
        title = "{The continued spectral and temporal evolution of RX J0720.4-3125}",
      journal = {\mnras},
         year = 2012,
        month = jun,
       volume = {423},
       number = {2},
        pages = {1194-1199},
          doi = {10.1111/j.1365-2966.2012.20946.x},
archivePrefix = {arXiv},
       eprint = {1203.3708},
 primaryClass = {astro-ph.HE},
       adsurl = {https://ui.adsabs.harvard.edu/abs/2012MNRAS.423.1194H}
}

@ARTICLE{Dessert2020,
       author = {{Dessert}, Christopher and {Foster}, Joshua W. and {Safdi}, Benjamin R.},
        title = "{Hard X-Ray Excess from the Magnificent Seven Neutron Stars}",
      journal = {\apj},
         year = 2020,
        month = nov,
       volume = {904},
       number = {1},
          eid = {42},
        pages = {42},
          doi = {10.3847/1538-4357/abb4ea},
       adsurl = {https://ui.adsabs.harvard.edu/abs/2020ApJ...904...42D}
}

@BOOK{Gonzaga2012,
   author = {{Gonzaga {et al.}}, S.},
    title = "{The DrizzlePac Handbook}",
booktitle = {The DrizzlePac Handbook, HST Data Handbook},
     year = 2012,
    month = jun,
   adsurl = {http://adsabs.harvard.edu/abs/2012drzp.book.....G}
}

@INPROCEEDINGS{Fruchter2010,
   author = {{Fruchter {et al.}}, A.~S.},
    title = "{BetaDrizzle: A Redesign of the MultiDrizzle Package}",
booktitle = {2010 Space Telescope Science Institute Calibration Workshop, p. 382-387},
     year = 2010,
    month = jul,
    pages = {382-387},
   adsurl = {http://adsabs.harvard.edu/abs/2010bdrz.conf..382F}
}

@ARTICLE{Pires2019,
       author = {{Pires}, A.~M. and {Schwope}, A.~D. and {Haberl}, F. and {Zavlin}, V.~E. and {Motch}, C. and {Zane}, S.},
        title = "{A deep XMM-Newton look on the thermally emitting isolated neutron star RX J1605.3+3249}",
      journal = {\aap},
         year = 2019,
        month = mar,
       volume = {623},
          eid = {A73},
        pages = {A73},
          doi = {10.1051/0004-6361/201834801},
archivePrefix = {arXiv},
       eprint = {1901.08533},
 primaryClass = {astro-ph.HE},
       adsurl = {https://ui.adsabs.harvard.edu/abs/2019A&A...623A..73P}
}

@ARTICLE{Kurpas2024,
       author = {{Kurpas}, J. and {Schwope}, A.~D. and {Pires}, A.~M. and {Haberl}, F.},
        title = "{Detection of pulsed X-ray emission from the isolated neutron star candidate eRASSU J131716.9-402647}",
      journal = {arXiv e-prints},
         year = 2024,
        month = jan,
          eid = {arXiv:2401.17290},
        pages = {arXiv:2401.17290},
          doi = {10.48550/arXiv.2401.17290},
archivePrefix = {arXiv},
       eprint = {2401.17290},
 primaryClass = {astro-ph.HE},
       adsurl = {https://ui.adsabs.harvard.edu/abs/2024arXiv240117290K}
}

@ARTICLE{Posselt2012,
       author = {{Posselt}, B. and {Arumugasamy}, P. and {Pavlov}, G.~G. and {Manchester}, R.~N. and {Shannon}, R.~M. and {Kargaltsev}, O.},
        title = "{XMM-Newton Observation of the Very Old Pulsar J0108-1431}",
      journal = {\apj},
         year = 2012,
        month = dec,
       volume = {761},
       number = {2},
          eid = {117},
        pages = {117},
          doi = {10.1088/0004-637X/761/2/117},
archivePrefix = {arXiv},
       eprint = {1210.7179},
 primaryClass = {astro-ph.HE},
       adsurl = {https://ui.adsabs.harvard.edu/abs/2012ApJ...761..117P}
}

@ARTICLE{Posselt2023a,
       author = {{Posselt}, B. and {Karastergiou}, A. and {Johnston}, S. and {Parthasarathy}, A. and {Oswald}, L.~S. and {Main}, R.~A. and {Basu}, A. and {Keith}, M.~J. and {Song}, X. and {Weltevrede}, P. and {Tiburzi}, C. and {Bailes}, M. and {Buchner}, S. and {Geyer}, M. and {Kramer}, M. and {Spiewak}, R. and {Krishnan}, V. Venkatraman},
        title = "{The Thousand Pulsar Array program on MeerKAT - IX. The time-averaged properties of the observed pulsar population}",
      journal = {\mnras},
         year = 2023,
        month = apr,
       volume = {520},
       number = {3},
        pages = {4582-4600},
          doi = {10.1093/mnras/stac3383},
archivePrefix = {arXiv},
       eprint = {2211.11849},
 primaryClass = {astro-ph.HE},
       adsurl = {https://ui.adsabs.harvard.edu/abs/2023MNRAS.520.4582P}
}

@ARTICLE{Posselt2018,
       author = {{Posselt}, B. and {Pavlov}, G.~G. and {Ertan}, {\"U}. and {{\c{C}}al{\i}{\c{s}}kan}, S. and {Luhman}, K.~L. and {Williams}, C.~C.},
        title = "{Discovery of Extended Infrared Emission around the Neutron Star RXJ0806.4-4123}",
      journal = {\apj},
         year = 2018,
        month = sep,
       volume = {865},
       number = {1},
          eid = {1},
        pages = {1},
          doi = {10.3847/1538-4357/aad6df},
       adsurl = {https://ui.adsabs.harvard.edu/abs/2018ApJ...865....1P}
}

@ARTICLE{Ertan2017,
       author = {{Ertan}, {\"U}. and {{\c{C}}al{\i}{\c{s}}kan}, {\c{S}}. and {Alpar}, M.~A.},
        title = "{Optical excess of dim isolated neutron stars}",
      journal = {\mnras},
         year = 2017,
        month = may,
       volume = {470},
       number = {1},
        pages = {1253-1258},
          doi = {10.1093/mnras/stx1310},
       adsurl = {https://ui.adsabs.harvard.edu/abs/2017MNRAS.470.1253E}
}

@INPROCEEDINGS{Pires2023,
       author = {{Mancini Pires}, Adriana and {Schwope}, Axel and {Kurpas}, Jan},
        title = "{Deep eROSITA observations of the magnificent seven isolated neutron stars}",
    booktitle = {Neutron Star Astrophysics at the Crossroads: Magnetars and the Multimessenger Revolution},
         year = 2023,
       editor = {{Troja}, Eleonora and {Baring}, Matthew G.},
       volume = {363},
        month = jan,
        pages = {288-292},
          doi = {10.1017/S1743921322000333},
archivePrefix = {arXiv},
       eprint = {2202.06793},
 primaryClass = {astro-ph.HE},
       adsurl = {https://ui.adsabs.harvard.edu/abs/2023IAUS..363..288M}
}

@ARTICLE{Posselt2014,
       author = {{Posselt}, B. and {Pavlov}, G.~G. and {Popov}, S. and {Wachter}, S.},
        title = "{Herschel and Spitzer Observations of Slowly Rotating, Nearby Isolated Neutron Stars}",
      journal = {\apjs},
         year = 2014,
        month = nov,
       volume = {215},
       number = {1},
          eid = {3},
        pages = {3},
          doi = {10.1088/0067-0049/215/1/3},
       adsurl = {https://ui.adsabs.harvard.edu/abs/2014ApJS..215....3P}
}

@ARTICLE{Kaplan2011,
       author = {{Kaplan}, D.~L. and {Kamble}, A. and {van Kerkwijk}, M.~H. and {Ho}, W.~C.~G.},
        title = "{New Optical/Ultraviolet Counterparts and the Spectral Energy Distributions of Nearby, Thermally Emitting, Isolated Neutron Stars}",
      journal = {\apj},
         year = 2011,
        month = aug,
       volume = {736},
       number = {2},
          eid = {117},
        pages = {117},
          doi = {10.1088/0004-637X/736/2/117},
       adsurl = {https://ui.adsabs.harvard.edu/abs/2011ApJ...736..117K}
}

@ARTICLE{Kaplan2009,
       author = {{Kaplan}, D.~L. and {van Kerkwijk}, M.~H.},
        title = "{Constraining the Spin-down of the Nearby Isolated Neutron Star RX J0806.4-4123, and Implications for the Population of Nearby Neutron Stars}",
      journal = {\apj},
         year = 2009,
        month = nov,
       volume = {705},
       number = {1},
        pages = {798-808},
          doi = {10.1088/0004-637X/705/1/798},
       adsurl = {https://ui.adsabs.harvard.edu/abs/2009ApJ...705..798K}
}

@ARTICLE{Kaplan2009b,
       author = {{Kaplan}, D.~L. and {van Kerkwijk}, M.~H.},
        title = "{Constraining the Spin-Down of the Nearby Isolated Neutron Star RX J2143.0+0654}",
      journal = {\apjl},
         year = 2009,
        month = feb,
       volume = {692},
       number = {1},
        pages = {L62-L66},
          doi = {10.1088/0004-637X/692/1/L62},
archivePrefix = {arXiv},
       eprint = {0901.4133},
 primaryClass = {astro-ph.HE},
       adsurl = {https://ui.adsabs.harvard.edu/abs/2009ApJ...692L..62K}
}

@ARTICLE{Kondratiev2009,
       author = {{Kondratiev}, V.~I. and {McLaughlin}, M.~A. and {Lorimer}, D.~R. and {Burgay}, M. and {Possenti}, A. and {Turolla}, R. and {Popov}, S.~B. and {Zane}, S.},
        title = "{New Limits on Radio Emission from X-ray Dim Isolated Neutron Stars}",
      journal = {\apj},
         year = 2009,
        month = sep,
       volume = {702},
       number = {1},
        pages = {692-706},
          doi = {10.1088/0004-637X/702/1/692},
       adsurl = {https://ui.adsabs.harvard.edu/abs/2009ApJ...702..692K}
}

@ARTICLE{Hare2021,
       author = {{Hare}, Jeremy and {Volkov}, Igor and {Pavlov}, George G. and {Kargaltsev}, Oleg and {Johnston}, Simon},
        title = "{Precise Timing and Phase-resolved Spectroscopy of the Young Pulsar J1617-5055 with NuSTAR}",
      journal = {\apj},
         year = 2021,
        month = dec,
       volume = {923},
       number = {2},
          eid = {249},
        pages = {249},
          doi = {10.3847/1538-4357/ac30e2},
archivePrefix = {arXiv},
       eprint = {2110.08077},
 primaryClass = {astro-ph.HE},
       adsurl = {https://ui.adsabs.harvard.edu/abs/2021ApJ...923..249H}
}

@ARTICLE{Vahdat2024,
       author = {{Vahdat}, Armin and {Posselt}, Bettina and {Pavlov}, George G. and {Weltevrede}, Patrick and {Santangelo}, Andrea and {Johnston}, Simon},
        title = "{Multiwavelength pulsations and surface temperature distribution in the middle-aged pulsar B1055-52}",
      journal = {arXiv e-prints},
         year = 2024,
        month = jan,
          eid = {arXiv:2401.12373},
        pages = {arXiv:2401.12373},
          doi = {10.48550/arXiv.2401.12373},
archivePrefix = {arXiv},
       eprint = {2401.12373},
 primaryClass = {astro-ph.HE},
       adsurl = {https://ui.adsabs.harvard.edu/abs/2024arXiv240112373V}
}

@ARTICLE{Pires2022,
       author = {{Pires}, A.~M. and {Motch}, C. and {Kurpas}, J. and {Schwope}, A.~D. and {Valdes}, F. and {Haberl}, F. and {Traulsen}, I. and {Tub{\'\i}n}, D. and {Becker}, W. and {Comparat}, J. and {Maitra}, C. and {Meisner}, A. and {Moustakas}, J. and {Salvato}, M.},
        title = "{XMM-Newton and SRG/eROSITA observations of the isolated neutron star candidate 4XMM J022141.5{\ensuremath{-}}735632}",
      journal = {\aap},
         year = 2022,
        month = oct,
       volume = {666},
          eid = {A148},
        pages = {A148},
          doi = {10.1051/0004-6361/202244514},
archivePrefix = {arXiv},
       eprint = {2208.07637},
 primaryClass = {astro-ph.HE},
       adsurl = {https://ui.adsabs.harvard.edu/abs/2022A&A...666A.148P}
}

@ARTICLE{Posselt2008,
       author = {{Posselt}, B. and {Popov}, S.~B. and {Haberl}, F. and {Tr{\"u}mper}, J. and {Turolla}, R. and {Neuh{\"a}user}, R.},
        title = "{The needle in the haystack: where to look for more isolated cooling neutron stars}",
      journal = {\aap},
         year = 2008,
        month = may,
       volume = {482},
       number = {2},
        pages = {617-629},
          doi = {10.1051/0004-6361:20078430},
archivePrefix = {arXiv},
       eprint = {0801.4567},
 primaryClass = {astro-ph},
       adsurl = {https://ui.adsabs.harvard.edu/abs/2008A&A...482..617P}
}

@ARTICLE{Haberl2006,
       author = {{Haberl}, F. and {Turolla}, R. and {de Vries}, C.~P. and {Zane}, S. and {Vink}, J. and {M{\'e}ndez}, M. and {Verbunt}, F.},
        title = "{Evidence for precession of the isolated neutron star <ASTROBJ>RX J0720.4-3125</ASTROBJ>}",
      journal = {\aap},
         year = 2006,
        month = may,
       volume = {451},
       number = {2},
        pages = {L17-L21},
          doi = {10.1051/0004-6361:20065093},
archivePrefix = {arXiv},
       eprint = {astro-ph/0603724},
 primaryClass = {astro-ph},
       adsurl = {https://ui.adsabs.harvard.edu/abs/2006A&A...451L..17H}
}

@ARTICLE{vanKerkwijk2007,
       author = {{van Kerkwijk}, Marten H. and {Kaplan}, David L. and {Pavlov}, George G. and {Mori}, Kaya},
        title = "{Spectral and Rotational Changes in the Isolated Neutron Star RX J0720.4-3125}",
      journal = {\apjl},
         year = 2007,
        month = apr,
       volume = {659},
       number = {2},
        pages = {L149-L152},
          doi = {10.1086/518030},
archivePrefix = {arXiv},
       eprint = {astro-ph/0703326},
 primaryClass = {astro-ph},
       adsurl = {https://ui.adsabs.harvard.edu/abs/2007ApJ...659L.149V}
}

@ARTICLE{vanKerkwijk2001,
       author = {{van Kerkwijk}, M.~H. and {Kulkarni}, S.~R.},
        title = "{An unusual H{\ensuremath{\alpha}} nebula around the nearby neutron star <ASTROBJ>RX J1856.5-3754</ASTROBJ>}",
      journal = {\aap},
         year = 2001,
        month = dec,
       volume = {380},
        pages = {221-237},
          doi = {10.1051/0004-6361:20011386},
       adsurl = {https://ui.adsabs.harvard.edu/abs/2001A&A...380..221V}
}

@ARTICLE{Braje2002,
       author = {{Braje}, Timothy M. and {Romani}, Roger W.},
        title = "{RX J1856-3754: Evidence for a Stiff Equation of State}",
      journal = {\apj},
         year = 2002,
        month = dec,
       volume = {580},
       number = {2},
        pages = {1043-1047},
          doi = {10.1086/343895},
archivePrefix = {arXiv},
       eprint = {astro-ph/0208069},
 primaryClass = {astro-ph},
       adsurl = {https://ui.adsabs.harvard.edu/abs/2002ApJ...580.1043B}
}

@ARTICLE{Pavlov1996,
       author = {{Pavlov}, G.~G. and {Zavlin}, V.~E. and {Truemper}, J. and {Neuhaeuser}, R.},
        title = "{Multiwavelength Observations of Isolated Neutron Stars as a Tool to Probe the Properties of their Surfaces}",
      journal = {\apjl},
         year = 1996,
        month = nov,
       volume = {472},
        pages = {L33},
          doi = {10.1086/310355},
archivePrefix = {arXiv},
       eprint = {astro-ph/9609097},
 primaryClass = {astro-ph},
       adsurl = {https://ui.adsabs.harvard.edu/abs/1996ApJ...472L..33P}
}

@MISC{Avila2019,
       author = {{Avila}, R.~J. and {Bohlin}, R. and {Hathi}, N. and {Lockwood}, S. and {Lim}, P.~L. and {De La Pena}, M.},
        title = "{SBC Absolute Flux Calibration}",
 howpublished = {Instrument Science Report ACS 2019-5, 13 pages},
         year = 2019,
        month = oct,
        pages = {5},
       adsurl = {https://ui.adsabs.harvard.edu/abs/2019acs..rept....5A}
}

@ARTICLE{Zampieri2001,
       author = {{Zampieri}, L. and {Campana}, S. and {Turolla}, R. and {Chieregato}, M. and {Falomo}, R. and {Fugazza}, D. and {Moretti}, A. and {Treves}, A.},
        title = "{1RXS J214303.7+065419/RBS 1774: A new Isolated Neutron Star candidate}",
      journal = {\aap},
         year = 2001,
        month = oct,
       volume = {378},
        pages = {L5-L9},
          doi = {10.1051/0004-6361:20011151},
archivePrefix = {arXiv},
       eprint = {astro-ph/0108456},
 primaryClass = {astro-ph},
       adsurl = {https://ui.adsabs.harvard.edu/abs/2001A&A...378L...5Z}
}

@ARTICLE{Zane2005,
       author = {{Zane}, S. and {Cropper}, M. and {Turolla}, R. and {Zampieri}, L. and {Chieregato}, M. and {Drake}, J.~J. and {Treves}, A.},
        title = "{XMM-Newton Detection of Pulsations and a Spectral Feature in the X-Ray Emission of the Isolated Neutron Star 1RXS J214303.7+065419/RBS 1774}",
      journal = {\apj},
         year = 2005,
        month = jul,
       volume = {627},
       number = {1},
        pages = {397-403},
          doi = {10.1086/430138},
archivePrefix = {arXiv},
       eprint = {astro-ph/0503239},
 primaryClass = {astro-ph},
       adsurl = {https://ui.adsabs.harvard.edu/abs/2005ApJ...627..397Z}
}

@ARTICLE{Cropper2007,
       author = {{Cropper}, Mark and {Zane}, Silvia and {Turolla}, Roberto and {Zampieri}, Luca and {Chieregato}, Matteo and {Drake}, Jeremy and {Treves}, Aldo},
        title = "{XMM-Newton observations of the isolated neutron star 1RXS J214303.7+065419/RBS1774}",
      journal = {\apss},
         year = 2007,
        month = apr,
       volume = {308},
       number = {1-4},
        pages = {161-166},
          doi = {10.1007/s10509-007-9377-z},
       adsurl = {https://ui.adsabs.harvard.edu/abs/2007Ap&SS.308..161C}
}

@ARTICLE{Schwope2009,
       author = {{Schwope}, A.~D. and {Erben}, T. and {Kohnert}, J. and {Lamer}, G. and {Steinmetz}, M. and {Strassmeier}, K. and {Zinnecker}, H. and {Bechtold}, J. and {Diolaiti}, E. and {Fontana}, A. and {Gallozzi}, S. and {Giallongo}, E. and {Ragazzoni}, R. and {de Santis}, C. and {Testa}, V.},
        title = "{The isolated neutron star RBS1774 revisited. Revised XMM-Newton X-ray parameters and an optical counterpart from deep LBT-observations}",
      journal = {\aap},
         year = 2009,
        month = may,
       volume = {499},
       number = {1},
        pages = {267-272},
          doi = {10.1051/0004-6361/200811041},
archivePrefix = {arXiv},
       eprint = {0902.4110},
 primaryClass = {astro-ph.GA},
       adsurl = {https://ui.adsabs.harvard.edu/abs/2009A&A...499..267S}
}

@ARTICLE{Bogdanov2024,
       author = {{Bogdanov}, Slavko and {Ho}, Wynn C.~G.},
        title = "{The ``Magnificent Seven'' X-Ray Isolated Neutron Stars Revisited. I. Improved Timing Solutions and Pulse Profile Analysis}",
      journal = {\apj},
         year = 2024,
        month = jul,
       volume = {969},
       number = {1},
          eid = {53},
        pages = {53},
          doi = {10.3847/1538-4357/ad452b},
archivePrefix = {arXiv},
       eprint = {2407.00275},
 primaryClass = {astro-ph.HE},
       adsurl = {https://ui.adsabs.harvard.edu/abs/2024ApJ...969...53B}
}

@ARTICLE{Zane2008,
       author = {{Zane}, S. and {Mignani}, R.~P. and {Turolla}, R. and {Treves}, A. and {Haberl}, F. and {Motch}, C. and {Zampieri}, L. and {Cropper}, M.},
        title = "{An Optical Counterpart Candidate for the Isolated Neutron Star RBS 1774}",
      journal = {\apj},
         year = 2008,
        month = jul,
       volume = {682},
       number = {1},
        pages = {487-491},
          doi = {10.1086/589539},
archivePrefix = {arXiv},
       eprint = {0804.4394},
 primaryClass = {astro-ph},
       adsurl = {https://ui.adsabs.harvard.edu/abs/2008ApJ...682..487Z}
}

@ARTICLE{Posselt2009,
       author = {{Posselt}, B. and {Neuh{\"a}user}, R. and {Haberl}, F.},
        title = "{Searching for substellar companions of young isolated neutron stars}",
      journal = {\aap},
         year = 2009,
        month = mar,
       volume = {496},
       number = {2},
        pages = {533-545},
          doi = {10.1051/0004-6361/200810156},
archivePrefix = {arXiv},
       eprint = {0811.0398},
 primaryClass = {astro-ph},
       adsurl = {https://ui.adsabs.harvard.edu/abs/2009A&A...496..533P}
}

@MISC{Pavlov2023,
       author = {{Pavlov}, George G. and {Posselt}, Bettina and {Kargaltsev}, Oleg Y.},
        title = "{The most puzzling UV-optical-NIR spectrum of an isolated neutron star: A disk or a magnetosphere?}",
 howpublished = {HST Proposal. Cycle 31, ID. \#17476},
         year = 2023,
        month = aug,
        pages = {17476},
       adsurl = {https://ui.adsabs.harvard.edu/abs/2023hst..prop17476P}
}

@ARTICLE{Edenhofer2024,
       author = {{Edenhofer}, Gordian and {Zucker}, Catherine and {Frank}, Philipp and {Saydjari}, Andrew K. and {Speagle}, Joshua S. and {Finkbeiner}, Douglas and {En{\ss}lin}, Torsten A.},
        title = "{A parsec-scale Galactic 3D dust map out to 1.25 kpc from the Sun}",
      journal = {\aap},
         year = 2024,
        month = may,
       volume = {685},
          eid = {A82},
        pages = {A82},
          doi = {10.1051/0004-6361/202347628},
archivePrefix = {arXiv},
       eprint = {2308.01295},
 primaryClass = {astro-ph.GA},
       adsurl = {https://ui.adsabs.harvard.edu/abs/2024A&A...685A..82E}
}

@ARTICLE{Pavlov1997,
       author = {{Pavlov}, G.~G. and {Welty}, A.~D. and {C{\'o}rdova}, F.~A.},
        title = "{Hubble Space Telescope Observations of the Middle-aged Pulsar 0656+14}",
      journal = {\apjl},
         year = 1997,
        month = nov,
       volume = {489},
       number = {1},
        pages = {L75-L78},
          doi = {10.1086/310968},
       adsurl = {https://ui.adsabs.harvard.edu/abs/1997ApJ...489L..75P}
}

@ARTICLE{Skelton2014,
       author = {{Skelton}, Rosalind E. and {Whitaker}, Katherine E. and {Momcheva}, Ivelina G. and {Brammer}, Gabriel B. and {van Dokkum}, Pieter G. and {Labb{\'e}}, Ivo and {Franx}, Marijn and {van der Wel}, Arjen and {Bezanson}, Rachel and {Da Cunha}, Elisabete and {Fumagalli}, Mattia and {F{\"o}rster Schreiber}, Natascha and {Kriek}, Mariska and {Leja}, Joel and {Lundgren}, Britt F. and {Magee}, Daniel and {Marchesini}, Danilo and {Maseda}, Michael V. and {Nelson}, Erica J. and {Oesch}, Pascal and {Pacifici}, Camilla and {Patel}, Shannon G. and {Price}, Sedona and {Rix}, Hans-Walter and {Tal}, Tomer and {Wake}, David A. and {Wuyts}, Stijn},
        title = "{3D-HST WFC3-selected Photometric Catalogs in the Five CANDELS/3D-HST Fields: Photometry, Photometric Redshifts, and Stellar Masses}",
      journal = {\apjs},
         year = 2014,
        month = oct,
       volume = {214},
       number = {2},
          eid = {24},
        pages = {24},
          doi = {10.1088/0067-0049/214/2/24},
archivePrefix = {arXiv},
       eprint = {1403.3689},
 primaryClass = {astro-ph.GA},
       adsurl = {https://ui.adsabs.harvard.edu/abs/2014ApJS..214...24S}
}

@ARTICLE{Foight2016,
       author = {{Foight}, Dillon R. and {G{\"u}ver}, Tolga and {{\"O}zel}, Feryal and {Slane}, Patrick O.},
        title = "{Probing X-Ray Absorption and Optical Extinction in the Interstellar Medium Using Chandra Observations of Supernova Remnants}",
      journal = {\apj},
         year = 2016,
        month = jul,
       volume = {826},
       number = {1},
          eid = {66},
        pages = {66},
          doi = {10.3847/0004-637X/826/1/66},
archivePrefix = {arXiv},
       eprint = {1504.07274},
 primaryClass = {astro-ph.HE},
       adsurl = {https://ui.adsabs.harvard.edu/abs/2016ApJ...826...66F}
}

@ARTICLE{Lallement2022,
       author = {{Lallement}, R. and {Vergely}, J.~L. and {Babusiaux}, C. and {Cox}, N.~L.~J.},
        title = "{Updated Gaia-2MASS 3D maps of Galactic interstellar dust}",
      journal = {\aap},
         year = 2022,
        month = may,
       volume = {661},
          eid = {A147},
        pages = {A147},
          doi = {10.1051/0004-6361/202142846},
archivePrefix = {arXiv},
       eprint = {2203.01627},
 primaryClass = {astro-ph.GA},
       adsurl = {https://ui.adsabs.harvard.edu/abs/2022A&A...661A.147L}
}

@ARTICLE{Vergely2022,
       author = {{Vergely}, J.~L. and {Lallement}, R. and {Cox}, N.~L.~J.},
        title = "{Three-dimensional extinction maps: Inverting inter-calibrated extinction catalogues}",
      journal = {\aap},
         year = 2022,
        month = aug,
       volume = {664},
          eid = {A174},
        pages = {A174},
          doi = {10.1051/0004-6361/202243319},
archivePrefix = {arXiv},
       eprint = {2205.09087},
 primaryClass = {astro-ph.GA},
       adsurl = {https://ui.adsabs.harvard.edu/abs/2022A&A...664A.174V}
}

@ARTICLE{Leike2020,
       author = {{Leike}, R.~H. and {Glatzle}, M. and {En{\ss}lin}, T.~A.},
        title = "{Resolving nearby dust clouds}",
      journal = {\aap},
         year = 2020,
        month = jul,
       volume = {639},
          eid = {A138},
        pages = {A138},
          doi = {10.1051/0004-6361/202038169},
archivePrefix = {arXiv},
       eprint = {2004.06732},
 primaryClass = {astro-ph.GA},
       adsurl = {https://ui.adsabs.harvard.edu/abs/2020A&A...639A.138L}
}

@ARTICLE{Wilms2000,
       author = {{Wilms}, J. and {Allen}, A. and {McCray}, R.},
        title = "{On the Absorption of X-Rays in the Interstellar Medium}",
      journal = {\apj},
         year = 2000,
        month = oct,
       volume = {542},
       number = {2},
        pages = {914-924},
          doi = {10.1086/317016},
archivePrefix = {arXiv},
       eprint = {astro-ph/0008425},
 primaryClass = {astro-ph},
       adsurl = {https://ui.adsabs.harvard.edu/abs/2000ApJ...542..914W}
}

@ARTICLE{Abramkin2025,
       author = {{Abramkin}, Vadim and {Pavlov}, George G. and {Shibanov}, Yuriy and {Posselt}, B. and {Kargaltsev}, Oleg},
        title = "{The middle-aged pulsar PSR J1741{\textendash}2054 and its bow-shock nebula in the far-ultraviolet}",
      journal = {\aap},
         year = 2025,
        month = apr,
       volume = {696},
          eid = {A121},
        pages = {A121},
          doi = {10.1051/0004-6361/202452211},
archivePrefix = {arXiv},
       eprint = {2503.10540},
 primaryClass = {astro-ph.HE},
       adsurl = {https://ui.adsabs.harvard.edu/abs/2025A&A...696A.121A}
}

@ARTICLE{Medin2007,
       author = {{Medin}, Zach and {Lai}, Dong},
        title = "{Condensed surfaces of magnetic neutron stars, thermal surface emission, and particle acceleration above pulsar polar caps}",
      journal = {\mnras},
         year = 2007,
        month = dec,
       volume = {382},
       number = {4},
        pages = {1833-1852},
          doi = {10.1111/j.1365-2966.2007.12492.x},
       adsurl = {https://ui.adsabs.harvard.edu/abs/2007MNRAS.382.1833M}
}

@ARTICLE{Perez-Azorin2006,
       author = {{P{\'e}rez-Azor{\'\i}n}, J.~F. and {Miralles}, J.~A. and {Pons}, J.~A.},
        title = "{Anisotropic thermal emission from magnetized neutron stars}",
      journal = {\aap},
         year = 2006,
        month = jun,
       volume = {451},
       number = {3},
        pages = {1009-1024},
          doi = {10.1051/0004-6361:20054403},
archivePrefix = {arXiv},
       eprint = {astro-ph/0510684},
 primaryClass = {astro-ph},
       adsurl = {https://ui.adsabs.harvard.edu/abs/2006A&A...451.1009P}
}

@ARTICLE{Burwitz2003,
       author = {{Burwitz}, V. and {Haberl}, F. and {Neuh{\"a}user}, R. and {Predehl}, P. and {Tr{\"u}mper}, J. and {Zavlin}, V.~E.},
        title = "{The thermal radiation of the isolated neutron star RX J1856.5-3754 observed with Chandra and XMM-Newton}",
      journal = {\aap},
         year = 2003,
        month = mar,
       volume = {399},
        pages = {1109-1114},
          doi = {10.1051/0004-6361:20021747},
archivePrefix = {arXiv},
       eprint = {astro-ph/0211536},
 primaryClass = {astro-ph},
       adsurl = {https://ui.adsabs.harvard.edu/abs/2003A&A...399.1109B}
}

@ARTICLE{Kaplan2003,
       author = {{Kaplan}, D.~L. and {van Kerkwijk}, M.~H. and {Marshall}, H.~L. and {Jacoby}, B.~A. and {Kulkarni}, S.~R. and {Frail}, D.~A.},
        title = "{The Nearby Neutron Star RX J0720.4-3125 from Radio to X-Rays}",
      journal = {\apj},
         year = 2003,
        month = jun,
       volume = {590},
       number = {2},
        pages = {1008-1019},
          doi = {10.1086/375052},
archivePrefix = {arXiv},
       eprint = {astro-ph/0303126},
 primaryClass = {astro-ph},
       adsurl = {https://ui.adsabs.harvard.edu/abs/2003ApJ...590.1008K}
}

@ARTICLE{Yakovlev2021,
       author = {{Yakovlev}, D.~G.},
        title = "{Two-blackbody portraits of radiation from magnetized neutron stars}",
      journal = {\mnras},
         year = 2021,
        month = sep,
       volume = {506},
       number = {3},
        pages = {4593-4602},
          doi = {10.1093/mnras/stab2077},
archivePrefix = {arXiv},
       eprint = {2107.10761},
 primaryClass = {astro-ph.HE},
       adsurl = {https://ui.adsabs.harvard.edu/abs/2021MNRAS.506.4593Y}
}

@ARTICLE{Greenstein1983,
       author = {{Greenstein}, G. and {Hartke}, G.~J.},
        title = "{Pulselike character of blackbody radiation from neutron stars.}",
      journal = {\apj},
         year = 1983,
        month = aug,
       volume = {271},
        pages = {283-293},
          doi = {10.1086/161195},
       adsurl = {https://ui.adsabs.harvard.edu/abs/1983ApJ...271..283G}
}

@ARTICLE{Arumugasamy2018,
       author = {{Arumugasamy}, Prakash and {Kargaltsev}, Oleg and {Posselt}, Bettina and {Pavlov}, George G. and {Hare}, Jeremy},
        title = "{Possible Phase-dependent Absorption Feature in the X-Ray Spectrum of the Middle-aged PSR J0659+1414}",
      journal = {\apj},
         year = 2018,
        month = dec,
       volume = {869},
       number = {2},
          eid = {97},
        pages = {97},
          doi = {10.3847/1538-4357/aaec69},
archivePrefix = {arXiv},
       eprint = {1810.11814},
 primaryClass = {astro-ph.HE},
       adsurl = {https://ui.adsabs.harvard.edu/abs/2018ApJ...869...97A}
}

@ARTICLE{Ho2007,
       author = {{Ho}, Wynn C.~G. and {Kaplan}, David L. and {Chang}, Philip and {van Adelsberg}, Matthew and {Potekhin}, Alexander Y.},
        title = "{Magnetic hydrogen atmosphere models and the neutron star RX J1856.5-3754}",
      journal = {\mnras},
         year = 2007,
        month = mar,
       volume = {375},
       number = {3},
        pages = {821-830},
          doi = {10.1111/j.1365-2966.2006.11376.x},
archivePrefix = {arXiv},
       eprint = {astro-ph/0612145},
 primaryClass = {astro-ph},
       adsurl = {https://ui.adsabs.harvard.edu/abs/2007MNRAS.375..821H}
}

@ARTICLE{Pons2009,
       author = {{Pons}, J.~A. and {Miralles}, J.~A. and {Geppert}, U.},
        title = "{Magneto-thermal evolution of neutron stars}",
      journal = {\aap},
         year = 2009,
        month = mar,
       volume = {496},
       number = {1},
        pages = {207-216},
          doi = {10.1051/0004-6361:200811229},
archivePrefix = {arXiv},
       eprint = {0812.3018},
 primaryClass = {astro-ph},
       adsurl = {https://ui.adsabs.harvard.edu/abs/2009A&A...496..207P}
}

@ARTICLE{Perez-Azorin2005,
       author = {{P{\'e}rez-Azor{\'\i}n}, J.~F. and {Miralles}, J.~A. and {Pons}, J.~A.},
        title = "{Thermal radiation from magnetic neutron star surfaces}",
      journal = {\aap},
         year = 2005,
        month = apr,
       volume = {433},
       number = {1},
        pages = {275-283},
          doi = {10.1051/0004-6361:20041612},
archivePrefix = {arXiv},
       eprint = {astro-ph/0410664},
 primaryClass = {astro-ph},
       adsurl = {https://ui.adsabs.harvard.edu/abs/2005A&A...433..275P}
}

@ARTICLE{Posselt2024,
       author = {{Posselt}, B. and {Pavlov}, G.~G. and {Ho}, W.~C.~G. and {Haberl}, F.},
        title = "{NICER Timing of the X-Ray Thermal Isolated Neutron Star RX J0806.4{\textendash}4123}",
      journal = {\apj},
         year = 2024,
        month = sep,
       volume = {972},
       number = {2},
          eid = {197},
        pages = {197},
          doi = {10.3847/1538-4357/ad5f8c},
archivePrefix = {arXiv},
       eprint = {2407.04337},
 primaryClass = {astro-ph.HE},
       adsurl = {https://ui.adsabs.harvard.edu/abs/2024ApJ...972..197P}
}

@ARTICLE{Green2019,
       author = {{Green}, Gregory M. and {Schlafly}, Edward and {Zucker}, Catherine and {Speagle}, Joshua S. and {Finkbeiner}, Douglas},
        title = "{A 3D Dust Map Based on Gaia, Pan-STARRS 1, and 2MASS}",
      journal = {\apj},
         year = 2019,
        month = dec,
       volume = {887},
       number = {1},
          eid = {93},
        pages = {93},
          doi = {10.3847/1538-4357/ab5362},
archivePrefix = {arXiv},
       eprint = {1905.02734},
 primaryClass = {astro-ph.GA},
       adsurl = {https://ui.adsabs.harvard.edu/abs/2019ApJ...887...93G}
}

@ARTICLE{Hare2024,
       author = {{Hare}, Jeremy and {Pavlov}, George G. and {Posselt}, Bettina and {Kargaltsev}, Oleg and {Temim}, Tea and {Chen}, Steven},
        title = "{Probing the Spectrum of the Magnetar 4U 0142+61 with JWST}",
      journal = {\apj},
         year = 2024,
        month = sep,
       volume = {972},
       number = {2},
          eid = {176},
        pages = {176},
          doi = {10.3847/1538-4357/ad5ce5},
archivePrefix = {arXiv},
       eprint = {2405.03947},
 primaryClass = {astro-ph.HE},
       adsurl = {https://ui.adsabs.harvard.edu/abs/2024ApJ...972..176H}
}

@ARTICLE{Kargaltsev2007,
       author = {{Kargaltsev}, Oleg and {Pavlov}, George},
        title = "{Ultraviolet emission from young and middle-aged pulsars: Connecting X-rays with the optical}",
      journal = {\apss},
         year = 2007,
        month = apr,
       volume = {308},
       number = {1-4},
        pages = {287-296},
          doi = {10.1007/s10509-007-9383-1},
archivePrefix = {arXiv},
       eprint = {astro-ph/0609656},
 primaryClass = {astro-ph},
       adsurl = {https://ui.adsabs.harvard.edu/abs/2007Ap&SS.308..287K}
}

@ARTICLE{vanAdelsberg2005,
       author = {{van Adelsberg}, Matthew and {Lai}, Dong and {Potekhin}, Alexander Y. and {Arras}, Phil},
        title = "{Radiation from Condensed Surface of Magnetic Neutron Stars}",
      journal = {\apj},
         year = 2005,
        month = aug,
       volume = {628},
       number = {2},
        pages = {902-913},
          doi = {10.1086/430871},
archivePrefix = {arXiv},
       eprint = {astro-ph/0406001},
 primaryClass = {astro-ph},
       adsurl = {https://ui.adsabs.harvard.edu/abs/2005ApJ...628..902V}
}

@ARTICLE{AstropyCollaboration2022,
       author = {{Astropy Collaboration} and {Price-Whelan}, Adrian M. and {Lim}, Pey Lian and {Earl}, Nicholas and {Starkman}, Nathaniel and {Bradley}, Larry and {Shupe}, David L. and {Patil}, Aarya A. and {Corrales}, Lia and {Brasseur}, C.~E. and {N{\"o}the}, Maximilian and {Donath}, Axel and {Tollerud}, Erik and {Morris}, Brett M. and {Ginsburg}, Adam and {Vaher}, Eero and {Weaver}, Benjamin A. and {Tocknell}, James and {Jamieson}, William and {van Kerkwijk}, Marten H. and {Robitaille}, Thomas P. and {Merry}, Bruce and {Bachetti}, Matteo and {G{\"u}nther}, H. Moritz and {Aldcroft}, Thomas L. and {Alvarado-Montes}, Jaime A. and {Archibald}, Anne M. and {B{\'o}di}, Attila and {Bapat}, Shreyas and {Barentsen}, Geert and {Baz{\'a}n}, Juanjo and {Biswas}, Manish and {Boquien}, M{\'e}d{\'e}ric and {Burke}, D.~J. and {Cara}, Daria and {Cara}, Mihai and {Conroy}, Kyle E. and {Conseil}, Simon and {Craig}, Matthew W. and {Cross}, Robert M. and {Cruz}, Kelle L. and {D'Eugenio}, Francesco and {Dencheva}, Nadia and {Devillepoix}, Hadrien A.~R. and {Dietrich}, J{\"o}rg P. and {Eigenbrot}, Arthur Davis and {Erben}, Thomas and {Ferreira}, Leonardo and {Foreman-Mackey}, Daniel and {Fox}, Ryan and {Freij}, Nabil and {Garg}, Suyog and {Geda}, Robel and {Glattly}, Lauren and {Gondhalekar}, Yash and {Gordon}, Karl D. and {Grant}, David and {Greenfield}, Perry and {Groener}, Austen M. and {Guest}, Steve and {Gurovich}, Sebastian and {Handberg}, Rasmus and {Hart}, Akeem and {Hatfield-Dodds}, Zac and {Homeier}, Derek and {Hosseinzadeh}, Griffin and {Jenness}, Tim and {Jones}, Craig K. and {Joseph}, Prajwel and {Kalmbach}, J. Bryce and {Karamehmetoglu}, Emir and {Ka{\l}uszy{\'n}ski}, Miko{\l}aj and {Kelley}, Michael S.~P. and {Kern}, Nicholas and {Kerzendorf}, Wolfgang E. and {Koch}, Eric W. and {Kulumani}, Shankar and {Lee}, Antony and {Ly}, Chun and {Ma}, Zhiyuan and {MacBride}, Conor and {Maljaars}, Jakob M. and {Muna}, Demitri and {Murphy}, N.~A. and {Norman}, Henrik and {O'Steen}, Richard and {Oman}, Kyle A. and {Pacifici}, Camilla and {Pascual}, Sergio and {Pascual-Granado}, J. and {Patil}, Rohit R. and {Perren}, Gabriel I. and {Pickering}, Timothy E. and {Rastogi}, Tanuj and {Roulston}, Benjamin R. and {Ryan}, Daniel F. and {Rykoff}, Eli S. and {Sabater}, Jose and {Sakurikar}, Parikshit and {Salgado}, Jes{\'u}s and {Sanghi}, Aniket and {Saunders}, Nicholas and {Savchenko}, Volodymyr and {Schwardt}, Ludwig and {Seifert-Eckert}, Michael and {Shih}, Albert Y. and {Jain}, Anany Shrey and {Shukla}, Gyanendra and {Sick}, Jonathan and {Simpson}, Chris and {Singanamalla}, Sudheesh and {Singer}, Leo P. and {Singhal}, Jaladh and {Sinha}, Manodeep and {Sip{\H{o}}cz}, Brigitta M. and {Spitler}, Lee R. and {Stansby}, David and {Streicher}, Ole and {{\v{S}}umak}, Jani and {Swinbank}, John D. and {Taranu}, Dan S. and {Tewary}, Nikita and {Tremblay}, Grant R. and {de Val-Borro}, Miguel and {Van Kooten}, Samuel J. and {Vasovi{\'c}}, Zlatan and {Verma}, Shresth and {de Miranda Cardoso}, Jos{\'e} Vin{\'\i}cius and {Williams}, Peter K.~G. and {Wilson}, Tom J. and {Winkel}, Benjamin and {Wood-Vasey}, W.~M. and {Xue}, Rui and {Yoachim}, Peter and {Zhang}, Chen and {Zonca}, Andrea and {Astropy Project Contributors}},
        title = "{The Astropy Project: Sustaining and Growing a Community-oriented Open-source Project and the Latest Major Release (v5.0) of the Core Package}",
      journal = {\apj},
         year = 2022,
        month = aug,
       volume = {935},
       number = {2},
          eid = {167},
        pages = {167},
          doi = {10.3847/1538-4357/ac7c74},
archivePrefix = {arXiv},
       eprint = {2206.14220},
 primaryClass = {astro-ph.IM},
       adsurl = {https://ui.adsabs.harvard.edu/abs/2022ApJ...935..167A}
}

@INPROCEEDINGS{Gabriel2004,
       author = {{Gabriel}, C. and {Denby}, M. and {Fyfe}, D.~J. and {Hoar}, J. and {Ibarra}, A. and {Ojero}, E. and {Osborne}, J. and {Saxton}, R.~D. and {Lammers}, U. and {Vacanti}, G.},
        title = "{The XMM-Newton SAS - Distributed Development and Maintenance of a Large Science Analysis System: A Critical Analysis}",
    booktitle = {Astronomical Data Analysis Software and Systems (ADASS) XIII},
         year = 2004,
       editor = {{Ochsenbein}, Francois and {Allen}, Mark G. and {Egret}, Daniel},
       series = {Astronomical Society of the Pacific Conference Series},
       volume = {314},
        month = jul,
        pages = {759},
       adsurl = {https://ui.adsabs.harvard.edu/abs/2004ASPC..314..759G}
}

@ARTICLE{Gordon2023,
       author = {{Gordon}, Karl D. and {Clayton}, Geoffrey C. and {Decleir}, Marjorie and {Fitzpatrick}, E.~L. and {Massa}, Derck and {Misselt}, Karl A. and {Tollerud}, Erik J.},
        title = "{One Relation for All Wavelengths: The Far-ultraviolet to Mid-infrared Milky Way Spectroscopic R(V)-dependent Dust Extinction Relationship}",
      journal = {\apj},
         year = 2023,
        month = jun,
       volume = {950},
       number = {2},
          eid = {86},
        pages = {86},
          doi = {10.3847/1538-4357/accb59},
archivePrefix = {arXiv},
       eprint = {2304.01991},
 primaryClass = {astro-ph.GA},
       adsurl = {https://ui.adsabs.harvard.edu/abs/2023ApJ...950...86G}
}

@ARTICLE{Geppert2006,
       author = {{Geppert}, U. and {K{\"u}ker}, M. and {Page}, D.},
        title = "{Temperature distribution in magnetized neutron star crusts. II. The effect of a strong toroidal component}",
      journal = {\aap},
         year = 2006,
        month = oct,
       volume = {457},
       number = {3},
        pages = {937-947},
          doi = {10.1051/0004-6361:20054696},
archivePrefix = {arXiv},
       eprint = {astro-ph/0512530},
 primaryClass = {astro-ph},
       adsurl = {https://ui.adsabs.harvard.edu/abs/2006A&A...457..937G}
}

\end{document}